\documentclass[manuscript]{acmart}

\AtBeginDocument{
  \providecommand\BibTeX{{%
    \normalfont B\kern-0.5em{\scshape i\kern-0.25em b}\kern-0.8em\TeX}}}

\AddToHook{env/quote/begin}{\small\color{black!50}}

\newcommand{\iquote}[1]{\textit{\textcolor{black!50}{#1}}}

\newcommand{\pid}[1]{{\fontfamily{cmss}\selectfont{\scriptsize{\textbf{\textcolor{black!50}{(#1)}}}}}}

\usepackage{multicol}
\usepackage{colortbl}
\usepackage{multirow}
\usepackage{hhline}

\newcolumntype{L}[1]{>{\raggedright\arraybackslash}p{#1}}

\begin{document}

\title{Mental Models of Meeting Goals: Supporting Intentionality in Meeting Technologies}

\author{Ava Elizabeth Scott}
\authornote{Both authors contributed equally to this research.}
\authornote{The work was done when the co-author was employed at Microsoft.}
\affiliation{%
  \institution{University College London}
  \city{London}
  \country{United Kingdom}
}
\email{ava.scott.20@ucl.ac.uk}

\author{Lev Tankelevitch}
\authornotemark[1]
\affiliation{%
  \institution{Microsoft Research}
  \city{Cambridge}
  \country{United Kingdom}
}
\email{lev.tankelevitch@microsoft.com}

\author{Sean Rintel}
\affiliation{%
  \institution{Microsoft Research}
  \city{Cambridge}
  \country{United Kingdom}}
\email{serintel@microsoft.com}

\renewcommand{\shortauthors}{Scott, Tankelevitch, and Rintel}

\begin{abstract}

Ineffective meetings due to unclear goals are major obstacles to productivity, yet support for intentionality is surprisingly scant in our meeting and allied workflow technologies. To design for intentionality, we need to understand workers’ attitudes and practices around goals. We interviewed 21 employees of a global technology company and identified contrasting mental models of meeting goals: meetings as a means to an end, and meetings as an end in themselves. We explore how these mental models impact how meeting goals arise, goal prioritization, obstacles to considering goals, and how lack of alignment around goals may create tension between organizers and attendees. We highlight the challenges in balancing preparation, constraining scope, and clear outcomes, with the need for intentional adaptability and discovery in meetings. Our findings have implications for designing systems which increase effectiveness in meetings by catalyzing intentionality and reducing tension in the organisation of meetings. 
\end{abstract}



 \begin{teaserfigure}
   \includegraphics[width=\textwidth]{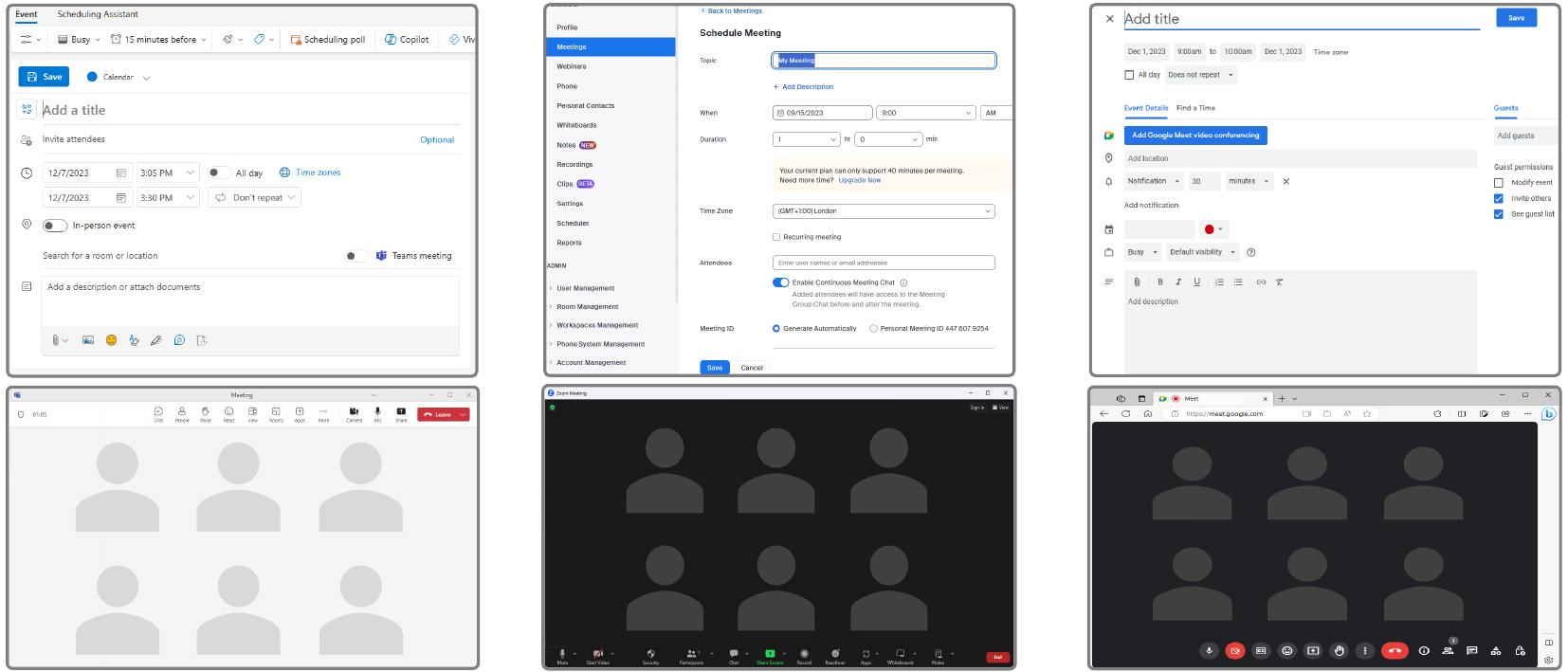}
   \caption{Goals are not explicitly featured in Microsoft, Zoom, or Google calendar invitation interfaces (top row) or videoconferencing interfaces (bottom row).}
   \Description{Six screenshots of interfaces are shown in two shows. The top row shows calendar invitations interfaces, and the bottom row shows videoconferencing interfaces. Neither include any mention of goals.}
   \label{fig:teaser}
 \end{teaserfigure}

\maketitle

\section{Introduction}
Meetings, as a collective means of social orientation and coordination, have become central to modern work~\cite{vanvree_formalisationofmeetings_2019, tropman2003making}. However, knowledge workers have been complaining about having too many meetings and ineffective meetings for decades~\cite{nixon_impact_1992, romano_meeting_2001,luong_meetings_2005,rogelberg_not_2006, leach_perceived_2009,bang_effectiveness_2010,cohen_meeting_2011,geimer_meetings_2015,rogelberg2018surprising}. In 2023, inefficient meetings and a lack of clear goals remained the top two obstacles to productivity according to Microsoft’s Work Trend Index~\cite{microsoft_work_2023}. 
In the COVID-19 pandemic, the phenomena of videoconferencing fatigue~\cite{doring_videoconference_2022} arose as a consequence of many work sectors being unprepared for conducting fully remote work, leading to (among other things) an over-reliance on video meetings without an intentional approach to their use~\cite{teevan2020new,bergmann_meeting_2023}. 
Unfortunately, the technologies that enable in-person, remote, or hybrid meetings often do not require users to reason about or explain \textit{why} a meeting is needed. Calendars fill up with meetings that lack explicit goals, and video meetings provide audio-visual canvases that do not in and of themselves promote goal-directed behavior. In this paper, we argue that to design and build technologies that can clarify and support intentionality in meetings, we first need to understand the predominant mental models of meeting goals, and the current practices used to achieve these purposes.

In the sections that follow, we outline Related Work (§\ref{relatedwork}) on meeting goals, agendas, meeting effectiveness, understanding meeting goals, and designing for intentionality. We then report our study Methods (§\ref{methods}), in which we interviewed 21 employees working across several different work areas in a global technology company about their understanding and practices around meeting goals. Our Findings (§\ref{findings}) detail two different mental models for meeting goals that stood out from the interviews: \textit{meetings as a means to an end} (where the goal is to achieve something), and \textit{meetings as an end in themselves} (where goals are seen as discussion topics, to connect with others, or block time). While the former model is associated with constrained scope, greater preparation, and clear meeting outcomes, the latter is associated with an elevated chance of discovery and adaptability to arising needs. 

Our Discussion (§\ref{discussion}) focuses on implications for the design of calendar and meeting systems to catalyze intentionality. There are many opportunities for systems to surface intentionality within different interfaces and time points across the meeting lifecycle. We explore how this implication might be accepted and treated through participant responses to an exploratory probe of adding a goals field to calendar invitation interfaces. We also speculate on how generative AI might further enhance intentionality in calendar and planning workflows, as well as on the potential for meeting interfaces that adapt to goal-directed behavior. Our Conclusion (§\ref{conclusion}) is that these findings and recommendations can inform a future where calendar and meeting systems facilitate fewer, better meetings, reducing interpersonal conflict and meeting fatigue, while increasing effectiveness. 

\section{Related Work}
\label{relatedwork}

Meeting science has explored characteristics that influence perceived meeting effectiveness~\cite{romano_meeting_2001, leach_perceived_2009, cohen_meeting_2011, geimer_meetings_2015, cutler_meeting_2021, bang_effectiveness_2010, nixon_impact_1992}, and types of meeting goals \cite{allen_understanding_2014, standaert_how_2021, romano_meeting_2001}. However, there is no generally agreed understanding or taxonomy of \textit{how} people think about and communicate meeting goals and the implications for collaboration.

Collaboration has long been a focus of research in Human-Computer Interaction (HCI), including extensive inquiries into the nature of coordination in Computer-Supported Cooperative Work (CSCW)~\cite{grudin_CSCWreview_1994,schmidt_cscwfoundations_2013}, explorations of video-mediated communication and media space systems~\cite{finn-sellen-wilbur_vmc_1997,harrison_mediaspace_2009}, and Group Support Systems~\cite{turoff_gss_1993,hiltz_gss_2015}. Some research has focused more directly on features to support intentionality, such as automated meeting agenda creation~\cite{raikundalia_autoagenda_1998}, automated meeting scheduling recommendations~\cite{haynes_autoscheduling_1997}, and the like. Other 
explorations have explored how technology-supported meetings progress or break down through a range of coordination phenomena including collaborative coupling and problem-solving~\cite{isenberg2012coupling}, social presence and co-presence~\cite{oh2018systematic}, and hybrid meeting experiences~\cite{neumayr2021hybrid, tang2023hybrid}. However, overall, HCI research has rarely developed from a holistic understanding of the goal-setting and communication issues in either experimental systems or commercial ecosystems. For example, the most common reason for failure in meetings relying on Group Support Systems (GSS) is the lack of clear meeting goals and pre-planning~\cite{de_vreede_how_2003}, suggesting this is an essential factor for meeting moderation and effectiveness, and yet holistic support for meeting goals is muted in reports about GSS systems (e.g. see~\cite{hiltz_gss_2015}). Across HCI, meeting goals tend to be mentioned loosely and in passing as part of an overall claim of the value of improved collaboration modalities, but most of this research has not sought to develop principled ways of supporting specifically goal-driven behaviour in meeting technologies.

Previous research has described meetings as being used as an all too convenient container for any conversation-based work action, perhaps with negative consequences~\cite{rogelberg2018surprising}. \citet{bergmann_meeting_2023} argue that the umbrella term of `meeting' fails to refer or specialize to the specific purposes and intentions of a collaborative encounter, and may thus obstruct both productivity and technology development. Reflecting this limited vocabulary, meeting science rarely operationalizes different kinds of meeting motivations~\cite{rogelberg2010employee, geimer_meetings_2015}, despite meeting goals being highlighted as critical for decades~\cite{strauss2004open}. 

\subsection{Meeting goals and agendas}

Goal-setting theory indicates that goals positively drive performance by increasing effort, focusing attention on goal-relevant activities, harnessing goal-relevant skills and knowledge, and encouraging persistence until goal achievement~\cite{latham_self-regulation_1991}. Meeting goals and agendas are closely related, but we suggest they are distinct. If goals set the destination for a meeting, then agendas outline the route. Agendas are therefore examples of implementation intentions: plans that “spell out the when, where, and how of goal striving in advance”; these have been consistently demonstrated to increase the likelihood of goal achievement ~\cite{gollwitzer_implementation_2006, sheeran_intention-behavior_2016}.

\subsection{Understanding meeting goals}

\citet{allen_understanding_2014} developed a taxonomy of meeting purposes based on a qualitative analysis of survey responses from an online panel of workers across organizations. Thirteen out of 16 purposes were framed as “discussions” (e.g., about an ongoing project, or a change in process), categorized as content-focused, and three were action-oriented (e.g., to educate, problem-solve, brainstorm), categorized as instrumental-focused. According to sensemaking theory, \citet{allen_understanding_2014} suggest that people retrospectively make sense of their meetings; employees attend meetings to resolve ambiguity about a topic and therefore primarily focus on the content aspect of meetings, rather than the specific goals the meeting is designed to achieve. In contrast, \citet{lopez-fresno_what_2022} found that participants %
commonly viewed meeting purposes in instrumental terms (e.g., to decide, plan, coordinate), and, less commonly, in social terms (e.g., to persuade, motivate, socialize), reflecting views of meetings as collaboration technology and as rituals (as per ~\cite{scott2015culture}). 

In the current work, we identify \textit{meetings as a means to an end} and \textit{meetings as ends in themselves} as two overarching mental models of meetings, which partly align with \citet{allen_understanding_2014}'s `instrumental-' and `content-' focused meeting categories (see §\ref{subsubsec:meetingsasmeans} and §\ref{subsubsec:meetingsasend}). More broadly, whereas \citet{allen_understanding_2014}'s taxonomy focuses on the content or topic of discussions occurring \textit{within} meetings (e.g., `to discuss a client's needs or wants'), our analysis abstracts away from content, and instead focuses on people's intentions and assumptions preceding the meeting, and their relationship to work outside the meeting. Our categorization also partly aligns with that of \citet{lopez-fresno_what_2022}, although their analysis does not capture people's common tendency to view meeting goals as a set of `discussion topics' not explicitly linked to external goals, which we suggest is an important marker of a lack of intentionality in collaboration (see §\ref{discussion}). 

Other research has also produced lists of meeting types, purposes, or goals ~\cite{soria_recurring_2022, standaert_empirical_2016, standaert_how_2021, romano_meeting_2001}. Differences in meeting goals also exist between one-off and recurring meetings. The latter may have implicit goals alongside an explicitly instrumental purpose to advance work (also analyzed in ~\cite{soria_recurring_2022}), including increasing group awareness, creating a space for collective thinking, and providing a mechanism for setting deadlines, bonding, and gaining visibility in an organization ~\cite{niemantsverdriet_recurring_2017}. 

The majority of this work focuses on defining a set of specific meeting goals according to topics or actions occurring within meetings (e.g., `gathering knowledge' \cite{soria_recurring_2022}, `make a decision' \cite{standaert_empirical_2016}, `solve a problem' \cite{romano_meeting_2001}), in contrast to our approach of abstracting away from content or actions, and focusing in depth on people's mental models, including their intentions and assumptions that contextualize their meetings.  

Ideally, meeting goals would pertain to all attendees who should work together towards their achievement. The clarity of meeting goals therefore directly influences the clarity of each attendee's role~\cite{geimer_meetings_2015}, and ultimately meeting effectiveness. Indeed, ~\citet{bang_effectiveness_2010} found that the clarity of goals in meeting agendas and goal-focused communication during meetings was associated with team effectiveness. Attendees, however, may not always have a consistent understanding of meeting goals. \citet{lopez-fresno_what_2022} found that stated meeting purposes often differed from those perceived by meeting attendees. \citet{allen_understanding_2014} suggest that differences may also exist between meeting organizers, who may have an instrumental lens on meetings, and attendees, who may focus more on the content of meetings. People’s hidden agendas or personal goals that they bring to meetings may also be in tension with the collective goals ~\cite{romano_meeting_2001, lopez-fresno_what_2022}. Our findings provide evidence for the role and implications of meeting goal clarity, perceptions of meeting goals, and hidden agendas (see §\ref{practicesobstacles} and §\ref{implications}). 

Outside of meeting science, research on communication and group behavior has also examined why people meet to converse. \citet{mcgrath1991time} proposed that groups engage in activities to support three functions: production, group well-being, and individual member support. Similarly, \citet{yeomans_conversational_2022} proposed that conversations are characterized by an informational (exchanging accurate information) and relational (building a relationship) dimension. Importantly, productivity and sociality are not in opposition but are interlinked. For example, `small talk’ establishes common ground ~\cite{holmes2003small} and interpersonal trust ~\cite{ma2019interpersonal} which are integral to collaborative work ~\cite{mcgrath1991time, yoerger2015participate}. Our findings presenting the two different mental models of meeting goals speak to these two dimensions (see §\ref{subsubsec:meetingsasmeans} and §\ref{subsubsec:meetingsasend}).

\subsection{Designing for Intentionality}
Whereas some work has focused on improving group understanding \cite{kim2016improving, nunamaker_electronic_1991} or detecting conversational content \cite{zhou2021role} to predict meeting effectiveness \cite{zhou_navigating_2023}, to our knowledge, no socio-technical research has been done to support intentionality around meeting goals. Outside of meeting goals specifically, recent work has explored the role of intentionality in remote and hybrid meetings \cite{bergmann_meeting_2023}, including for social conversational transitions \cite{gonzalez_diaz_making_2022}, use of parallel chat \cite{sarkar_promise_2021}, the deployment of attention and multitasking \cite{cao2021large, kuzminykh_classification_2020}, and meeting participation \cite{samrose_meetingcoach_2021}. Our work is grounded in these findings, which show that intentionality may be implicitly present in some meeting contexts (e.g., in in-person meetings, the use of conversational transitions to conduct pre- and post- meeting talk \cite{gonzalez_diaz_making_2022}, or the use of eye gaze to signal attention \cite{kuzminykh_classification_2020}), but that this is not facilitated in videoconferencing interfaces. Accordingly, as we also demonstrate here, this lack of designing for intentionality may lead to misunderstandings and frustrated collaboration (e.g., as occurs with meeting multitasking \cite{cao2021large}), as well as lost productivity.     

HCI research has also explored the use of goal-setting, tracking, and reflection in the workplace \cite{meyer_enabling_2021, kocielnik_reflecting_2018} and in collaborative projects \cite{jackson_encouraging_2016}. Our design envisioning in §\ref{discussion} is grounded in this work, which demonstrates that purposeful self-reflection at work can increase self-awareness around one's daily and higher-level goals, improve work habits (including around management and organization), and change perspectives about work \cite{meyer_enabling_2021, kocielnik_reflecting_2018}. As a pre-requisite for intentionality, a line of research has explored designing for reflection (reviewed in \cite{bentvelzen_revisiting_2022, baumer_reflective_2015, baumer_reviewing_2014}). Our ideas in §\ref{discussion} build on this work, which discusses the value of self-reflective prompts, goal-setting, and slowing down to reflect \cite{bentvelzen_revisiting_2022, reicherts_make_2020}.    

\subsection{Summary}
Taken together, prior work has recognized the importance of meeting goals for meeting and organizational effectiveness, and proposed categorizations of meeting goals, but little work has examined meeting goals themselves: how people think about them, how they arise and are communicated or obstructed across the meeting lifecycle, and the effects this has on work and productivity. Similarly, there is limited work on how goal-directed behavior can be augmented and improved in planning, calendar, and meeting technologies. Thus the contributions of this paper are as follows:
\begin{itemize}
    \item Fundamental findings about social-technical issues around meeting goals: mental models, how goals arise, obstacles to goal behavior, and functional and affective implications.
    \item Design implications for catalyzing intentionality in calendar invitations and meeting technologies, from simple interventions to speculations on the uses of generative AI. 
\end{itemize}

\section{Methods}
\label{methods}
\subsection{Participants}
Following ethics authorization\footnote{Ethics authorization was provided by Microsoft Research's Institutional Review Board (IORG0008066, IRB00009672).}, we recruited a purposive sample ~\cite{etikan2016comparison} of 21 participants working across six different work areas within a global technology company. The sample was drawn from a company list of prior research participants who had indicated interest in future studies. We selected for a diversity of backgrounds, including gender, age, location, work area, seniority, hybrid work status, and managerial status (\autoref{tab:participants}; for more participant details, see Supplementary Materials).  Participants were consented before their interviews and offered a gift voucher for their participation.

\begin{table}
\small
  \caption{Participant demographics.}
  \label{tab:participants}
  \begin{tabular}{L{0.3\linewidth}  L{0.4\linewidth}  L{0.2\linewidth}}
    \toprule
    \textbf{Dimension}&\textbf{Sub-dimensions}&\textbf{Participants}\\
    \midrule
    Gender* & Male & 10\\
    & Female & 11\\
    \hline
Location & United States & 13  \\
& United Kingdom & 3                            \\
& Ireland & 2                                   \\
& Australia & 1                                 \\
& Colombia & 1                                  \\
& Denmark & 1                                   \\
\hline
Age & 18-29 & 2\\
 & 30-44 & 8                         \\
& 45-59 & 8                         \\
& 60 and over & 2                   \\
& Prefer not to say & 2             \\
\hline
Work area & Administration & 3 \\
& Customer Support & 2                          \\
& Product Development & 8                       \\
& Research & 2                                  \\
& Sales & 3                                     \\
& Technical and Facilities & 2                  \\
\hline
Seniority & Principal & 6 \\
& Senior & 12                               \\
& Early Career & 3                          \\
\hline
Managerial status & Manages a team & 8 \\
& Individual contributor & 13   \\    
\bottomrule
\end{tabular}
\newline *For gender, no participants identified as non-binary or declined to answer
\end{table}

\subsection{Interview Protocol and Analysis }

After online consent, participants completed an onboarding survey of demographics and work/meeting attitudes and behaviors. Microsoft Teams video meetings were then scheduled for recorded semi-structured interviews lasting between 30 and 70 minutes. Participants were guided through a short online survey as a prompt to start thinking about meeting goals and help recall, and the remaining time focused on discussion of the mental models and practices surrounding meeting goals, agendas, and outcomes. (See Supplementary Materials for details). 

Interviews were automatically transcribed by Microsoft Teams, and then further cleaned by the first author. Two authors coded six interviews using open coding ~\cite{strauss2004open}. This involved re-watching the interview recordings, reading the transcripts, and mapping out key themes on a virtual whiteboard for each interview. We focused on the interview responses as accounts of members’ methods for accomplishing social behaviour~\cite{heritage2013garfinkel}. This is important because to design technologies for meeting goal clarity, we must understand where accounts for meeting goals reveal assumptions and choice points, and how new technologies will themselves become part of future goal accounts. 

Through iterative discussions between all authors, an initial, hierarchical codebook was developed, with themes such as `Definitions of Goals', `Personality' and `Obstacles to thinking about meeting purpose'. Using the qualitative analysis tool, MAXQDA Plus 2022, the first author coded the remaining 15 interviews according to this code book. The authors dived into these high-level codes to find precise code quotes, e.g., specific examples of definitions of goals, personality effects, and obstacles to setting goals~\cite{braun2006using}. Using MAXQDA's creative coding feature, these codes were categorized and sorted into sub-codes, e.g., ``Goals as a getting something out of the meeting'', and ``Lack of time as an obstacle''. 

\subsection{Limitations}

In common with exploratory formative qualitative work, we recognize that our sample has several limitations. Our participant cohort consisted of employees from just one global technology firm, which has shaped their norms, practices, and available technologies. Further, although our sample contains a cross-section in terms of gender, age, work area, seniority, and managerial status, there may be patterns of aggregrated experiences or intersections with unique experiences that we have missed. For example, seventy percent of our participants were from the US, yet meeting practices clearly differ across cultures \cite{kohler_meetings_2015,lehmann-willenbrock_team-meeting_2017,van_eerde_meetings_2015}. For example, research suggests that meetings in the US tend to be short and goal-oriented, whereas meetings in the UK tend to be longer, and focus more on relationship-building, and meetings in Japan tend to focus on presentations and group consensus-building%
. Thus, our findings may not fully generalize to other cultures and future research needs to address both cultural variety and the effects of multi-cultural teams. Further, participants self-selected into our study based on participation in prior meeting research, and thus potentially were more thoughtful or concerned about meetings than a broader cohort might be. Moreover, meeting organizers may be biased in evaluating their own meetings \cite{cohen_meeting_2011}, and we had no method for directly identifying and accounting for bias. 

\section{Findings}
\label{findings}

\begin{figure*}[h]
\centering
  \includegraphics[width=\textwidth]{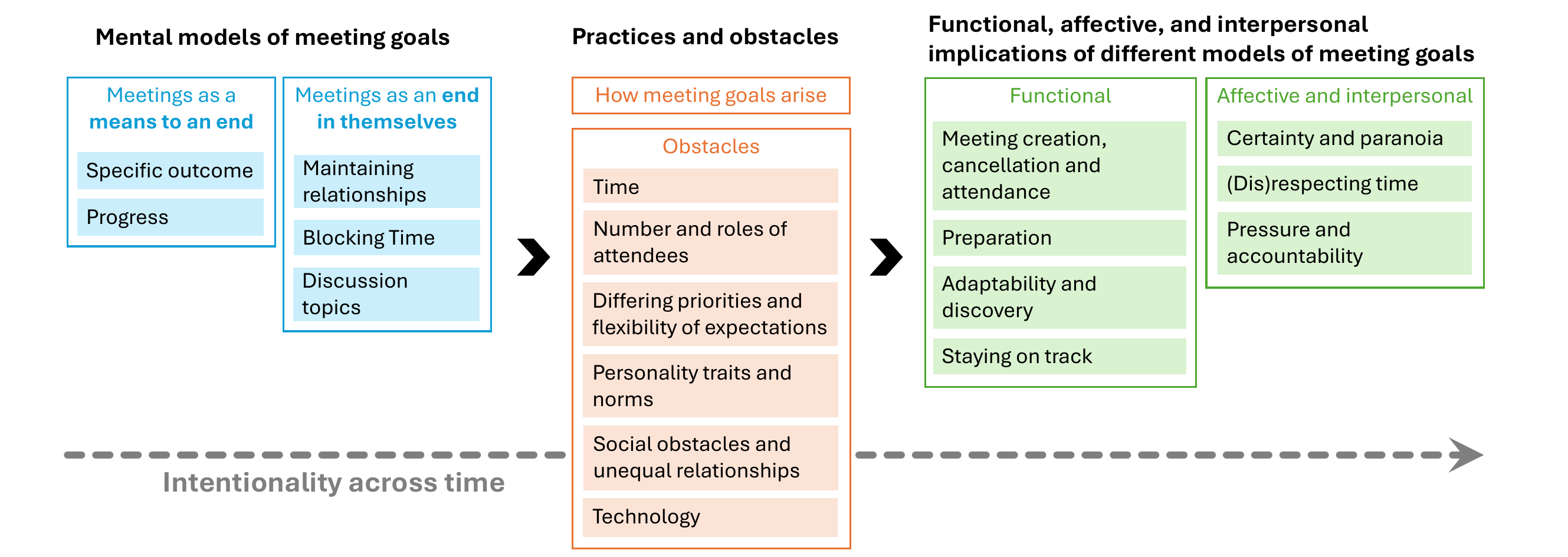}
  \caption{Summary of findings: our findings centered around mental models of meeting goals; current practices and obstacles to considering meeting goals; and the functional, affective, and interpersonal implications of these mental models, practices, and obstacles on meeting intentionality.}
  \Description{This image depicts a flow diagram with three main components, illustrating the progression from 'Mental models of meeting goals' through 'Practices and obstacles' to 'Functional, affective, and interpersonal implications of different models of meeting goals'. The first component, 'Mental models of meeting goals', is divided into 'Meetings as a means to an end', with subcategories 'Specific outcome' and 'Progress', and 'Meetings as an end in themselves', with subcategories 'Maintaining relationships', 'Blocking Time', and 'Discussion topics'. The second component, 'Practices and obstacles', contains 'How meeting goals arise' and a list of obstacles such as 'Time', 'Number and roles of attendees', 'Differing priorities and flexibility of expectations', 'Personality traits and norms', 'Social obstacles and unequal relationships', and 'Technology'. The third component details the 'Functional' implications like 'Meeting creation, cancellation and attendance', 'Preparation', 'Adaptability and discovery', and 'Staying on track', as well as 'Affective and interpersonal' implications like 'Certainty and paranoia', '(Dis)respecting time', 'Pressure and accountability'. The arrow labeled 'Intentionality across time' at the bottom signifies the direction of the flow and suggests that intentionality is present across all of the components.}
  \label{fig:findings}
\end{figure*}

Our findings are presented in three sections, summarized in \autoref{fig:findings}.
The first describes two overarching mental models of meeting goals; \textit{meetings as a means to an end}, and \textit{meetings as ends in themselves}. This section captures how the participants conceptualize meetings as mediating intentionality, or as directly representing an intention, respectively. We then report on current practices in setting and communicating meeting goals, including how meeting goals arise, and the obstacles to considering, communicating, and understanding meeting goals. This section shows how intentionality can be catalyzed by certain practices, and obstructed by a lack of time, as well as interpersonal and technological factors. Finally, we report on the functional and affective implications of these mental models, as perceived by our participants. This section presents the impact of the described mental models and practices on people's experiences of meetings and their varying levels of intentionality.

Intentionality can travel through the meeting lifecycle, from its conceptualization to its execution. However, this intentionality can be catalyzed or obstructed at different moments in this journey.  Throughout the findings, we highlight where issues of intentionality around meetings crop up. Our findings are supported by excerpts from the interviews, and interpreted alongside previous literature where relevant. For context, we include participants' work area and managerial status (i.e., individual contributor [IC] or manager) for each excerpt, formatted as such: \pid{P1 - Technical and Facilities, IC}. For details on each participant's gender, work area, managerial status, and seniority, see Supplementary Materials.

\subsection{Mental Models of Meeting Goals}

Two overarching mental models of the purpose of meetings stood out from our interviews. In the mental model of \textit{meetings as a means to an end}, participants view meetings as achieving outcomes that are expressly in aid of broader work. Meeting goals are defined and are specific, knowable, communicable, and achieved or not. By contrast, in the mental model of \textit{meetings as ends in themselves}, participants view meetings as time for connection or discussion. Meeting goals have more forms in this model. Goals may be known topics for discussion underpinned by an implicit assumption about how they aid broader work. Goals may be not defined at all, or at times even considered antithetical to the accomplishment of connection. Goals may not be pre-defined, but seen as discoverable and adaptable within the meeting. The goal may be blocking time itself as a placeholder for known or unknown work.  

These mental models are not mutually exclusive, they may be co-present within one individual or within one meeting, change across meetings, and be different for one-off and recurring meetings. This is because meeting goals can be relevant at the level of the meeting itself or at the level of the individual attendee. 

Of course, whatever mental model a participant may hold, meetings themselves are gestalt phenomena. Participants' mental models are just one thread in a larger tapestry of group actions and cognition, woven from a combination of engagement with the technological affordances used to set up and enact the meeting, communication about and in the meeting, and the meeting's place in organizational history and culture ~\cite{scott2015five}. 

\subsubsection{Meetings as a means to an end}\label{subsubsec:meetingsasmeans}
\label{meetings_as_means}
In this model, meeting goals have a dual nature as internal to the meeting and related to external work.   Intentionality is catalyzed by the meeting, with the goals defined before the meeting, achieved within the meeting, and informing work beyond the meeting. This can be achieved either through the delineation of a clear task that can be completely achieved within the meeting but is treated as meaningful within some broader context, or through discussion that clearly progresses work beyond the meeting. Thus, meetings are treated as a \textit{collaboration technology}~\cite{scott2015five} which enables people to achieve \textit{specific outcomes}, or to make \textit{progress} towards specific outcomes (i.e., they are \textit{instrumental} to achieving an outcome \cite{allen_understanding_2014}).    

\paragraph{The goal is a specific outcome}
\label{goal_is_outcome}
The core of this aspect of the mental model is an answerable question of purpose.  Such questions are proposed as having a direct answer where there will be an accountable change of state, e.g., a solution, a choice, completing a task, or explaining tools or processes \pid{P5, P2, P17}. The necessary conditions for achieving the change of state in the meeting exist before the meeting and will imply clear consequences after the meeting. For example, P3 \pid{Administration, Manager} explained that in a meeting to determine how a budget will be spent, the decisions made will impact work after the meeting.

\begin{quote}
    \iquote{``…What are we trying to accomplish here? What problem are we trying to solve?''} \pid{P12 - Product Development, IC} 
\end{quote}
\begin{quote}
\iquote{``...it's a kind of like a question you need to be able to answer by the end of the meeting, right? We will align or land on, um, pick a design or pick a plan or pick a date, you know.''}  \pid{P5 - Product Development, IC} 
\end{quote}

In this model, meeting goals are treated as knowable and communicable; if a meeting doesn't have a clear goal, then invited participants may \iquote{``not [be] sure why we’re having the meeting''} \pid{P1 - Technical and Facilities, IC}.  This creates a sense of obligation to provide a purpose \pid{P1 as an attendee; P13, P12 as meeting organizers}. This obligation may be part of the invitation process \pid{P13} and part of starting the meeting itself \pid{P12}. Starting the meeting with explicit articulation of intention reinforces the rationale for the meeting and establishes among attendees that this intention should be achieved by the end of the meeting. Indeed, meeting goal clarity and goal-focused discussions are related to team effectiveness \cite{bang_effectiveness_2010}.

\begin{quote}
\iquote{``If I'm calling a meeting, I will put either an agenda or why I've called the meeting in the invitation itself so that people know.''}  \pid{P13 - Research, IC}
\end{quote}
\begin{quote}
\iquote{``When I'm leading a meeting, I will always start with; this is why we're meeting.''}  \pid{P12 - Product Development, IC}
\end{quote}

Since this model involves knowing about an accountable transition in the state of the work, an effective meeting is one in which the change of state will be clear, as evidenced by P5:

\begin{quote}\iquote{``An action is we will brainstorm, right, and a goal could be, we landed on an idea. So maybe you did a brainstorm, but you didn't land on an idea.''} \pid{P5 - Product Development, IC}
\end{quote}

On some occasions meetings may end up being `more’ effective than expected because they accelerate a change of state \pid{P2}. Hence, attendees are explicitly aware of specific goals and whether they were achieved in the meeting. 
\begin{quote}
\iquote{``We didn’t expect to make the decision right then and there, but then it went so well that we ended up making the decision.''} \pid{P2 - Research, IC} 
\end{quote}

\paragraph{The goal is progress}
\label{goal_is_progress}

While it is possible to achieve certain goals by the end of the meeting, multiple participants described meeting outcomes that represent a step towards achieving a goal, or \textit{making progress}. When describing a first-time meeting organized by a customer, P16 described the outcome of a meeting as facilitating progress towards a goal.
\begin{quote}
    \iquote{``That is an outcome that's not gonna necessarily solve the goal in one step, but it's an outcome that's going to help us on that step''} \pid{P16 - Product Development, Manager}
\end{quote}

As the above quote suggests, whereas some meetings have social `goals' that are not tied to an explicit outcome (see §\ref{goal_is_relationships}), connecting to others for the first time as part of a broader project, or when informing work beyond the meeting, represents making progress towards a specific outcome. For example, a meeting can fulfill the intention of meeting clients, research subjects or other teams for the first time to establish a relationship or the potential for a relationship \pid{P7}. Many projects and collaborations start with a kick-off meeting, allowing relevant team members to come together and set up their intention for the work outside the meeting. 

\begin{quote}
    \iquote{``Our team goal was to explore whether or not it would be valuable to have a relationship with this other team.''} \pid{P7 - Product Development, IC} 
\end{quote}
\begin{quote}
\iquote{``...we don't have a lot of high touch with the individuals that have been invited. And so this is the first time that I'll be discussing this project with them.''} \pid{P12 - Product Development, IC}
\end{quote}
The intention of using meetings to make progress towards a specific outcome or as part of a larger work context particularly applies to recurring meetings set up to create \iquote{``rhythm through a big, complicated project''} and to \iquote{``keep momentum going''} \pid{P13 - Research, IC}. These regular meetings keep collaborators up-to-date on quickly changing projects \pid{P13}, and ensure that any blocks in progress are reviewed \pid{P6}.  Furthermore, these recurring meetings can demonstrate the reason for their own existence, by facilitating change until the broader intention is met. P16, a manager in product development, said a recurring meeting will only continue as long as the team is still required to work on a particular project.  

\begin{quote}
   \iquote{``…to keep momentum going … because things change so quickly, it’s important to like have a regular rhythm through a big, complicated project.''} \pid{P13 - Research, IC} 
\end{quote}
\begin{quote}
 \iquote{``We have an operational ongoing need to address to review changes to our road map plan to […] review where people are blocked on the most critical projects…''} \pid{P6 - Product Development, Manager}  
\end{quote}

While the overall intention behind a set of recurring meetings may be justified by an ongoing project, in practice attendees may find them problematic when any given instance is unlikely to achieve a change of state. A set of recurring meetings can stretch over a long time; in this period, the original intention may be lost, become irrelevant, or may not be best achieved through a regular meeting \cite{niemantsverdriet_recurring_2017}.

\begin{quote}
    \iquote{``I don't care for a lot of the recurring meetings just to have time to talk about things … because there's no purpose, it tends to get into what I just call group therapy, where everybody feels like they need to share something.''} \pid{P18 - Customer Support, IC} 
\end{quote}
\begin{quote}
\iquote{``I often find that recurring meetings just turned into a waste of time because there's no set agenda… One-off meetings are definitely more decision-driven, like there is an express purpose, like there is something that has arisen, and we're trying to address it.''} \pid{P15 - Customer Support, Manager} 
\end{quote}

\subsubsection{Meetings as an end in themselves}\label{subsubsec:meetingsasend}

In this mental model, the goal of a meeting is inextricable from the meeting itself. Meeting goals are not explicitly linked to specific outcomes, changes of state, or work outside the meeting. The act of meeting itself is the purpose, allowing the maintenance of relationships, general discussion of topics, or blocking of time. 

\paragraph{The goal is maintaining relationships}
\label{goal_is_relationships}
Seven participants \pid{P4, P6, P10, P14, P16, P18, P21} expressed the idea that the goal of the meeting was to be seen by and to connect with others, with varying levels of intentionality. After the first time (as noted above in §\ref{goal_is_outcome}), meeting regularly can be seen as a way to maintain a trusting relationship with new collaborators \pid{P16 - Product Development, Manager} and with teammates \pid{P7 - Product Development, IC}. In this way, meetings are an end in themselves because meeting organizers do not see any other way of achieving this abstract goal of connection \pid{P7}. While using meetings to maintain relationships and trust could be considered a means to an end, there is no specific outcome or change of state that will result from any specific instance—not unlike a ritual~\cite{scott2015five, lopez-fresno_what_2022, niemantsverdriet_recurring_2017}. Arguably, attending these meetings may become a form of presenteeism; P6 \pid{Product Development, Manager} mentioned attending meetings to support a teammate, or to get \iquote{``facetime''} with a senior colleague (see also \cite{niemantsverdriet_recurring_2017}). However, this does not have to be viewed cynically – there really are times when \textit{being there} matters \pid{P6, P14}. 
\begin{quote}
\iquote{``When I was a team leader, sometimes the only goal for me wasn't really so much to get something out of that kind of a meeting, other than making sure my team felt connected with each other because I think that's super important.''} \pid{P7 - Product Development, IC} 
\end{quote}

\begin{quote}
\iquote{``… if the customer puts a call out there it's always good to kind of for me to turn up and wave and kind of `Yes, we're here, we're keen, we're eager to help', rather than sometimes just do it blandly over email or a Teams site chat.'' }\pid{P14 - Product Development, IC} 
\end{quote}

\paragraph{The goal is discussing topics}
\label{goal_is_topics}
Beyond maintaining relationships, discussion topics can be treated as goals in themselves. The most diffuse version of this is to avoid explicit agendas and goals, on the assumption that openness and sharing are intrinsically valuable, as P21 expresses: 
\begin{quote}
         \iquote{``I would say that you have meetings where you don't have a defined agenda or you don't have a specific goal, you know, in mind, right? Your goal is basically really to be open and just share information.''} \pid{P21 - Administration, Manager}
\end{quote}

Many participants expressed a more focused treatment of discussion topics, such that the meeting goal is to cover certain topics \pid{P16} or get through the standing agenda for a recurring meeting \pid{P1}. The topics and their discussion were not explicitly linked to a goal outside the meeting, though they may be implicitly linked in the participant's mind. They articulate what they are discussing, but not \textit{why} they are discussing it—the intention is unclear. As a result, the conflation of discussion topics and meeting goals could obscure meeting intentionality, and the purpose of the meeting beyond the meeting itself. Similarly, \citet{allen_understanding_2014} distinguish between \textit{instrumental} and \textit{content}-focused meeting purposes, with the latter potentially reflecting participants' attempt to reduce organizational ambiguity about a specific topic (i.e., meetings as \textit{sensemaking}~\cite{scott2015five}), rather than view meetings as a collaboration tool to achieve external goals (as discussed above).

\begin{quote}
    \iquote{``…just in the meeting chat the day before or even sometimes an hour before and just be like “OK, guys today, like we're going to talk on these topics or we're going to cover these things.''} \pid{P16 - Product Development, Manager} 
\end{quote}
\begin{quote}
\iquote{``If it's a recurring meeting... they always have an agenda and, you know, the goal is just to get through the agenda.''} \pid{P1 - Technical and Facilities, IC}

\end{quote}

\paragraph{The goal is blocking time}

\label{goal_is_timeblock}
An implication of digital calendar systems is that meeting events can block time in peoples' schedules. While there is an understanding that this time will be used for work, the nature of that work, or its purpose in the broader project, is often undefined or perhaps undefinable at the point the meeting is set (especially common for recurring meetings) \pid{P16}. Often, these meetings are set with an expectation that they can be canceled, which is a product of the intense competition to grab people’s time in extremely busy calendars \pid{P10} (which, in a classic Catch-22, they also exacerbate).  Hence, blocking time for a meeting creates a vacuum for a future intention, but what that intention is and how it will be achieved remains entirely ambiguous.

\begin{quote}
\iquote{``…we need to get time on the calendar for everyone to prioritize getting together to talk about whatever topics there are at hand. And so, this is how we're going to do that, right? We're gonna just create this block.''} \pid{P16 - Product Development, Manager}
 \end{quote}

\begin{quote}
\iquote{``The reason they're recurring is that it's easier to get it on your calendar now for … a few months out than it would be for me come July 1st and say `Oh, I need to talk to you on July 10th.' Nobody's gonna be available, right?''} \pid{P10 - Sales, IC}

\end{quote}

The second effect of competition for time is that as the number of meetings increase, time for focused work decreases, and people become skeptical that work items will get done. This, in turn, leads to the blocking of time for meetings in which the organizer is actually booking attendees’ time on task, with the added requirement that the organizer will also have to spend that time supervising the completion of that task. P4 described booking time to supervise a manager's completion of a task, while P13 described blocking time for a project stakeholder to complete a task. As well as representing a problem for workflow and management, this also represents skepticism about the effectiveness of asynchronous work in a time-pressured context.
\begin{quote}
    \iquote{``If this was with certain management... I would need to, like, book 2 hours with them in order for them to open it up and look at it, and I babysit them while they do the task.''} \pid{P4 - Sales, IC}
\end{quote}
\begin{quote}
\iquote{``I need you to fix these descriptions and I will sit here while you do that.''} \pid{P13 - Research, IC} 
\end{quote}

\subsection{Practices and Obstacles}\label{practicesobstacles}

While the previous section explores two different mental models of meeting goals, this section explores how the participants describe their current practices in setting and communicating intentionality in meetings, if at all. We also describe the perceived obstacles to the process of specifying, sharing, and clarifying meeting goals. 

\subsubsection{How meeting goals arise}\label{howgoalsarise}
Meeting goals may be explicit, implicit, or emergent. The most obvious place for explicit goals is in the subject or description of a meeting invitation \pid{8 participants}, or in an agenda \pid{12 participants}, although these may include \iquote{``the general topic, but [...] no goals''} \pid{P15 - Customer Support, Manager}. Other participants communicate goals in chat or email close to the meeting time, including immediately beforehand or even as it starts \pid{8 participants}. Recurring meetings with well-defined work projects may also articulate goals for the next meetings \pid{P9} (as in \cite{niemantsverdriet_recurring_2017}). 
\begin{quote}
\iquote{``So the recurring meeting is probably getting an update on where we were from the last meeting […] and we have clear goals for, you know, the next meeting, that kind of thing.''} \pid{P9 - Product Development, IC}
\end{quote}

Meeting goals may not be explicitly articulated but can be \iquote{``obvious by context''} when the team or the work has a well-established context \pid{P6 - Product Development, Manager}. This contextual knowledge may come from experience, especially for recurring meetings with familiar people and domain expertise \pid{P4}. These meetings address a recurring and highly familiar intention, and so can be addressed without being explicitly acknowledged.

\begin{quote} 
\iquote{``I have an account team meeting every Monday at 3pm, and that’s a recurring meeting and it’s happened what feels like for my entire life, and so I look at that and I’m like oh, I know exactly what the format is, I know what the plan is, I know what I have to do to contribute to it.''} \pid{P4 - Sales, IC} 

\end{quote}

Some participants described intentionally organising meetings with loose structure, resembling a form of \iquote{``planned flexibility''} \pid{P20}. P20 thought this was particularly important for weekly team meetings, while P7 felt it was crucial for meetings about wellbeing, where the context is specified but particular goals are not, to ensure the openness needed to make the meeting valuable.
\begin{quote}

\iquote{``…I do have meetings where it's intentional that things are not very specifically defined because then you can go in the direction you need. So it's flexibility, but planned flexibility…''} \pid{P20 - Administration, Manager}  
\end{quote}
\begin{quote}
    \iquote{``People just come and talk to share if they have problems with their feelings and we help each other out. So that meeting, for instance, has no goal and it shouldn't have a goal…''} \pid{P7 - Product Development, IC}
\end{quote}

\subsubsection{Obstacles to Setting and Understanding Meeting Goals}\label{obstacles}

Participants reported certain obstacles to setting and communicating meeting goals, including the amount of time they have, the number and role of attendees at the meeting, personality factors, differing priorities of attendees, hidden agendas, and issues in meeting technologies.

\paragraph{Time}

Participants commonly saw a lack of time as an obstacle to considering meeting goals in general \pid{P5} or due to the urgency of an ad-hoc, impromptu meeting \pid{P3}.  

\begin{quote}
   \iquote{ ``I recognize […] feeling like I don't have time to plan and stuff like that.''} \pid{P5 - Product Development, IC} 
\end{quote}
\begin{quote}
\iquote{``There could be a meeting that's happening to reach an urgent decision, and so we wouldn't necessarily have time for planning etcetera.''} \pid{P3 - Administration, Manager} 
\end{quote}

Time between meetings may be either too short to propose meeting goals \pid{P1}, or, ironically, too long \pid{P16}.   

\begin{quote}
\iquote{``[W]e had like a stand-up sort of every day. There, it was very hard to set up an agenda every time…''}\pid{P1 - Technical and Facilities, IC} 
\end{quote}
\begin{quote}
    \iquote{``I send it two weeks before and I'm just like OK, I'm not even thinking about this until the night before.'' }\pid{P16 - Product Development, Manager} 
\end{quote}

If \iquote{``people are busy''} \pid{P14 - Product Development, IC} or they have \iquote{``a lot of [meetings] after the other''} \pid{P9 - Product Development, IC}, they may not have time to look at the meeting invite or the pre-read material. If the context is not clear from current work or relationship, one-off meetings created with short notice offer little time for attendees to think beforehand, requiring time to clarify goals as the meeting starts \pid{P3}. As a result, the meeting can be derailed just by trying to clarify the intentionality behind it, which can be intimidating for new joiners \pid{P6}.
\begin{quote}

\iquote{``…how much of that meeting can I derail to build that context versus do I need to be having those pre-discussions?''} \pid{P6 - Product Development, Manager} 
\end{quote}
\begin{quote}
\iquote{``…if it's just a very quick, can you join now? Then I'll just ask in the meeting if it's not clear.''} \pid{P3 - Administration, Manager}
\end{quote}

\paragraph{Number and roles of attendees}

If people see time as a limiting factor to considering meeting goals, it suggests that \textit{setting and communicating meeting goals can take up significant time}. The more people in attendance at the meeting, the more time it can take to clarify and communicate meeting intentionality.

Large \iquote{``costly''} meetings may require specific goals to be communicated \pid{P20 - Administration, Manager}, but large numbers of people with different roles may also obstruct the setting of very specific goals \pid{P10, P1}. As a meeting scales in size, meeting organizers do not have time to think about, or communicate, how the intentionality behind a meeting aligns with each attendee's role and perspective, particularly when attendees may lack the prior context necessary to make sense of the meeting. In some cases, meeting organizers may not know exactly who is attending before the meeting begins, e.g., in a Sales meeting with potential customers \pid{P10 - Sales, IC}. This does not imply that meeting goals must be set for each individual attendee, but that the specificity and clarity of meeting goals enables all attendees to understand their purpose for attending and, if necessary, prepare accordingly (as evidenced in §\ref{implications} below). Large meetings with diverse roles make this more challenging. 

\begin{quote}
\iquote{``You impact a lot of people if it's not done properly. It's a very costly, costly meeting. So to me, I feel a lot of pressure in making things the most efficient possible and making sure that I hit the goal.''} \pid{P20 - Administration, Manager}
\end{quote}
\begin{quote}
\iquote{``…with a lot of people, [goals] should be as […] high level and general as possible because you don't want people just sitting there while people are discussing a bunch of things they don't care about.''} \pid{P1 - Technical and Facilities, IC}

\end{quote}

On the other hand, organizers may set up smaller, regular meetings, which require less formal planning compared to large meetings \pid{P20}. Participants found that a large number of attendees can inhibit clear meeting outcomes, such as coming to a decision \pid{P19}. For P19 \pid{Product Development, Manager}, the \iquote{``sheer amount of voices''} makes large meetings inefficient for achieving agreement and goals.

\begin{quote}
\iquote{``[With a] much smaller group, much more regular frequency […] it's gonna be a lot less formal.''} \pid{P20 - Administration, Manager}
\end{quote}
\begin{quote}
\iquote{``I will try not to [invite] more than you know, 5-6 people because then it becomes very gruelling... the larger meeting is mostly about communicating, which again you can do async... If you wanna have a discussion and agreement and goals and things like that, it's better to do things in smaller audiences.''} \pid{P19 - Product Development, Manager} 
\end{quote}

\paragraph{Personality traits and norms}

Certain people may be more or less likely to set goals due to a more spontaneous, relaxed style \pid{P4}. On the other hand, those who are planning-oriented may not only want goals but also clear sub-goals in a linear order to feel that a meeting is valuable \pid{P7}. 
\begin{quote}

\iquote{``I'm very bad at including an agenda or anything in the meetings and I feel like that's my personality... I'm more comfortable to like, fly by the seat of my pants.''} \pid{P4 - Sales, IC}
\end{quote}
\begin{quote}

\iquote{``I need things to check off along the way, so I am not as great with ambiguity as some people are because I think `A + B= C'  – Not `A + B, oh wait, here comes C, ohh, there's E... Wait, we've still got to get to C...'''} \pid{P7 - Product Development, IC}
\end{quote}

Individual people's approaches to intentionality influence, and are influenced by, those of others. Managers are particularly able to establish a precedent \pid{P5}. For those in customer-facing roles, the meeting goals may be set according to the customers' preferences \pid{P14}.  Hence, the more people attending the meeting, the more difficult it can be to know how to be intentional in a way that will meet their varying expectations.
\begin{quote}

\iquote{``If my manager doesn’t do it, doesn’t expect me to do it, like we both know it be nice to do, but if it’s not like that [...] social weight on forcing you to do it.''}  \pid{P5 - Product Development, IC} 
\end{quote}
\begin{quote}
\iquote{``Depending on...how a customer interacts, it could be quite fluid. Some other customers would like to be a bit more rigid, so I could probably bring up, say, an ADO board in terms of tracking action items. Other customers don't necessarily want that, prefer more sort of flexible approach.''}  \pid{P14 - Product Development, IC}
\end{quote}

\paragraph{Differing priorities and flexibility of expectations}

The more attendees there are, the more likely they are to have different priorities, such as how to approach a particular project  \pid{P10 - Sales, IC}. A participant who worked in Sales had noticed that customers cancel meetings if the agenda is not explicitly aligned with their priorities that day  \pid{P4 - Sales, IC}.  Differences in expectations around flexibility can cause tensions. If a very clear goal or agenda is set, attendees may expect a meeting to cover all goals and not deviate \pid{P11 - Sales, IC}. Hence, organizers may not provide clear goals, to flexibly accommodate a range of different intentions.

\begin{quote}
    \iquote{``…my technical team is trying to solution and I’m like, pump the brakes, we’re not there yet, right? […] So that can be challenging because they’re their engineers and they wanna fix it….''}  \pid{P10 - Sales, IC}
\end{quote}

\begin{quote}
 \iquote{``It could be a double-edged sword cause you can't deviate from that topic. [… A] preconception of the meeting […] is fine in terms of preparation, but then it also limits you…''}  \pid{P11 - Sales, IC}

\end{quote}

\paragraph{Hidden agendas and unequal relationships}
 
A mismatch between an individual's actual intention for a meeting, and the explicitly communicated goal, may be used surreptitiously  \pid{P4}. To prevent resistance, organizers may limit goal communication in the invitation, waiting instead until the meeting starts  \pid{P15}. P4 and P15 are in Sales and Customer Support, respectively, so have to navigate the unequal relationship when organising meetings in which they are service providers.

 \begin{quote}
     \iquote{``I'm not going to blatantly say, `Hey, my goal for this call is to get you to schedule meetings and give me more information', but I will pepper that into the conversation to try to pull it out of him.'' } \pid{P4 - Sales, IC} 
\end{quote}
\begin{quote}
\iquote{``Maybe I won't say exactly in the invite, like `I'm concerned that we're behind on this', right? I won't actually say that, but I will in the meeting, preface it by like, `Hey, I wanted to get us together because I see that we're behind and I want to figure out how we can fix this.'''}  \pid{P15 - Customer Support, Manager}
 \end{quote}

In other instances, participants were invited to meetings where the goals were unclear. Sometimes, participants avoided clarifying the goals with an organizer, as they were concerned about being perceived as impolite  \pid{P2 - Research, IC}, critical  \pid{P4, P13, P18} or ignorant  \pid{P15 - Customer Support, Manager}. The perceived risks of offending the organizer are higher in an ambiguous or unequal relationship, meaning people are more likely to clarify the goals of a meeting \textit{if} they have an established relationship with the organizer. 

\begin{quote} 
\iquote{``This meeting's coming from an external customer and partner, so it's not something we can control. It’s one of those difficult situations where you would like to go back and go, `Could you put an agenda together, please?' […] But this tends not to work that well on that side of the relationship.''}  \pid{P4 - Sales, IC}
\end{quote}
\begin{quote}
\iquote{``I think it's probably that highly based on the attendees and the existing relationship I have with them because I don't want to offend them if I decline their meeting or if I'm a jerk about, hey, I'm not gonna come unless you tell me what the agenda is, alright?''}  \pid{P18 - Customer Support, IC}
\end{quote}

If a goal is communicated, this does not mean that attendees will understand or agree with it.  Organizers felt frustrated when attendees had different interpretations of the goal \pid{P4} or pushed back on the goal, particularly \pid{P21}. P5 \pid{Product Development, IC} reported having to \iquote{``fight to be taken into account''}, and have his perspective as a UX designer heard within a business strategy meeting.
\begin{quote}
    \iquote{``Sometimes you set a goal and everyone kind of interprets it differently, and so that can be super frustrating.''}  \pid{P4 - Sales, IC}
\end{quote}
\begin{quote}
    \iquote{``I feel like there is a lot of push back because [...] there is a fear of actually stating goals and working toward those goals''}  \pid{P21 - Administration, Manager}
\end{quote}

\paragraph{Technology}

In addition to calendar anxiety, scheduling meetings may obstruct thinking about meeting goals  \pid{P19}, especially if there is time pressure to join the next meeting  \pid{P4}.  Meeting organizers have to navigate multiple calendars and prior commitments and prioritize between them, leading to a high mental load during scheduling. The scheduling process is complex and time-consuming due to busy schedules, resulting in meetings booked far in advance  \pid{P13 - Research, IC} (see also \cite{sun_rhythm_2023}). A long time between scheduling a meeting and actually having it could end up diluting the intentionality behind the meeting, as priorities and goals shift and change.
\begin{quote}
   \iquote{ ``If I'm going to the scheduling window […] I'm mostly focused on the times, right? What, where are there any holes for me to put this on and not so caring anymore about the goals.''}  \pid{P19 - Product Development, Manager} 
 \end{quote}
\begin{quote}
\iquote{``I'm jumping to the next [meeting], so I need to quickly throw it on, which is why I don't fill out an agenda or anything.''}  \pid{P4 - Sales, IC}
\end{quote}

If an attendee is not part of pre-meetings or included in email threads or chats, they can miss critical context  \pid{P6}. The use of multiple platforms for scheduling meetings, communicating about the goals, and hosting the meetings, leads to disjointed information  \pid{P16, P18}. Hence, technological systems supposedly designed to support meetings and wider collaboration can actually lead to the fragmentation and obscuring of intentionality. 

\begin{quote}
    \iquote{``I think because I'm going to have these conversations separately in pre-meetings as opposed to spelling it out in the meeting itself, there could be some stuff lost in translation….''} \pid{P6 - Product Development, Manager}
 \end{quote}
 
\begin{quote}
\iquote{``I've shared it in the Outlook invite, but as soon as someone joins in Teams, none of that context, […] other than the title of the meeting, appears for me.[…] Meeting notes are in Teams or SharePoint, OneNote or Loop or whatever.''}  \pid{P16 - Product Development, Manager}
 \end{quote}

\subsection{Functional, Affective, and Interpersonal Implications}\label{implications}

Participants described how the goals, or lack of goals, for a meeting, can have implications for functional aspects of that meeting, as well as affective and interpersonal implications for organizers and attendees. 

\subsubsection{Functional Implications}\label{functionalimplications}

Goals, or a lack of them, can impact processes of creating, cancelling and attending meetings, as well as the scope of the discussion and preparation for the meeting.

\paragraph{Meeting creation, cancellation, and attendance}

Many of our participants talked about having too many meetings. In this context, people must prioritize between different meetings and often try to reclaim time back from meetings wherever they can, by cancelling meetings without purpose (as discussed in §\ref{goal_is_timeblock}). When a meeting is seen as a means to an end, the end-goal can be used as a prioritizing factor when deciding whether to hold a meeting, or whether asynchronous tools could work better to achieve this intention  \pid{P19}.

\begin{quote}
\iquote{``For the discussion, we have Azure DevOps. We have Teams for when we need to integrate. We have tools like Loop in order to collaborate and write things concurrently. Those are the ones that are useful to do asynchronously.''} \pid{P19 - Product Development, Manager} 
\end{quote}

Six participants  \pid{P3, P8, P10, P13, P18, P19} mentioned that they were less likely to attend a meeting without a clear goal. All participants reported having inquired with a meeting organizer to clarify the intention behind the meeting.

\begin{quote}
\iquote{``I always try to find out what is the purpose of the meeting. So sometimes based on that information, I'm going to not attend the meeting.''}  \pid{P8 - Customer Support, IC} 

\end{quote}

While not attending a meeting to reclaim time can be helpful for the invitee, it may not be beneficial for those who see the meeting as an end in itself, such as to connect with a customer.

\paragraph{Adaptability and discovery}

If the meeting is seen as primarily a place for discussion and ad-hoc problem-solving, organizers may think that adapting to the needs of the attendees is more important than stating a clear goal. People who work with customers often prioritize this, so they can align and adjust to their customers’ needs and desires at that given time. 
\begin{quote}
    \iquote{``I also want to make it clear that I'm here for [the customer] and we can discuss whatever [the customer] wants.''} \pid{P4 - Sales, IC}
\end{quote}

Team meetings may also benefit from prioritizing flexibility and adaptability over setting clear goals (or balancing ``order and openness'' \cite{niemantsverdriet_recurring_2017}). In particular, recurring meetings can be seen as a time and place for people to get together on immediate problems, whatever they may be. A meeting that can pivot and address issues that are \iquote{``top of mind''} for the attendees can be more reactive to \iquote{``pressing''} needs \pid{P20}.   

\begin{quote}
\iquote{``You should be able to just let go of something that was planned, to just go with something more pressing for the team at the time.''}  \pid{P20 - Administration, Manager} 
\end{quote}

For both internal meetings and meetings with external customers, there was an understanding that allowing the meeting discussion to progress freely could lead to the discovery of unexpected ideas or opportunities. Describing her role as the ``CIA of the company'', P4 \pid{Sales, IC} noted that a relaxed approach to meetings can help her to gather new information from customers. P3 \pid{Administration, Manager} explicitly described the trade-off between discovery and achieving goals.

\begin{quote}
    \iquote{``If it feels a little bit more off-the-cuff and loosey-goosey, customers tend to share more as opposed to constraining themselves to just the topics at hand.''}  \pid{P4 - Sales, IC}
\end{quote}

\begin{quote}
    \iquote{``You’re not necessarily going to achieve your goals and that's OK, because actually you participate in a much richer discussion for something else… Occasionally conversations can go off on tangents, but sometimes that might be a good thing.''} \pid{P3 - Administration, Manager}
\end{quote}

\paragraph{Staying on track}

While open discussion can lead to the exploration and discovery of new ideas, participants perceived that this must be balanced against the risk of getting side-tracked. While P4 preferred organising off-the-cuff meetings, she found that unclear goals can lead to a meeting getting off track. P9 thought that while not all meetings require an agenda, he expects there to be at least some \iquote{``guardrails''} for the conversation.  This tension was recognized in organizers and attendees alike, and in participants across product development \pid{P1}, sales \pid{P4, P9}, and administration \pid{P20, P21}. 

\begin{quote}
    \iquote{``We have a standard meeting right now, that's every other week and I feel like the goals are very vague and oftentimes we get on and the conversation is completely sidetracked by what's top of mind for them.''} \pid{P4 - Sales, IC}
\end{quote}

 \begin{quote}
     \iquote{``If there is no agenda… I guess you need to kind of be vigilant on staying on topic, right?''} \pid{P1 - Technical and Facilities, IC} 
 \end{quote}

When someone sees a meeting as a means to an end, they may perceive open discussions as an inefficient way to achieve these ends. P21's role in administration and strategy includes advising on and running meetings for others; she noted that talking alone may not be enough to achieve outcomes.

\begin{quote}
 \iquote{ ```We'll talk about it' doesn't always generate formidable, actionable work.''} \pid{P21 - Administration, Manager} 
\end{quote}

Some individuals see it as their responsibility or opportunity to make the most of the time demarcated by the meeting if there isn’t a clearly stated goal or agenda. They may use the time to \iquote{``get the floor to get all [their] stuff figured out''} \pid{P15 - Customer Support, Manager}, or to \iquote{``push and nudge the meeting''} even if they’re not the facilitator \pid{P21 - Administration, Manager}. 

Using the meeting time to achieve one’s own intentions can be perceived as domineering by other attendees, and was perceived to happen more often when there is no overall goal for the meeting.  
\begin{quote}
\iquote{``There are instances where some people, because they love to speak, love to talk, they go on and on and forgetting that we've run out of time. That generally happens within meetings where the agenda is not clear or, if there is and that's clear, people are just not following it.''} \pid{P2 - Research, IC} 

\end{quote}

\paragraph{Preparation}

If the expected goal of a meeting is communicated, it allows attendees to prepare for the meeting, to ensure those goals are achieved.  This preparation can be as simple as getting into the right \iquote{``mindset''} for a meeting \pid{P11 - Sales, IC}, or \iquote{``just turn [their] brain on the right channel'' }\pid{P13 - Research, IC}.

Some people, such as P18, reflected that he has to \iquote{``mentally prepare''} for meetings that lack explicit goals. 
\begin{quote}
\iquote{``When there is no agenda, for these a lot of these recurring meetings… I've prepared for the chaos of rambling about topics and participating in that context.''} \pid{P18 - Customer Support, IC} 
\end{quote}

Clear goals can also allow more intensive preparation, like gathering certain information or forming an opinion \pid{P12, P13, P14, P15.}

 \begin{quote}

\iquote{``It tells me how to show up. Is this a brainstorm or is it a share facts meeting? If it's brainstorm, then I might not prepare anything or…yeah, I might not prepare as much as if it's `bring your information'''} \pid{P13 - Research, IC}
 \end{quote}
\begin{quote}   
\iquote{``If I were creating an agenda for this, I would mark those out in bulleted points ahead of time.. and try and give everyone at least 48 hours to formulate an opinion or get the information or materials together so that our meeting runs smoother.''} \pid{P12 - Product Development, IC} 

 \end{quote}

A lack of clear goals or purpose for the meeting can lead to a lack of preparation for that meeting. P14 described how if certain attendees are \iquote{``not prepared for it, then … it can destroy the outcome of the meeting.''} \pid{P14 - Product Development, IC}
 
While customers may treat a meeting as a time for open consultation, P14 described how customers can become \iquote{``a bit despondent''} when he doesn’t have the right information to hand. 
\begin{quote}

\iquote{ ``If you had told me beforehand, I would have got the information for you. Now I'm in a position where all I can say is `I don't know, I will have to find the information for you' which kills the meeting because you can't proceed.''} \pid{P14 - Product Development, IC} 
\end{quote}

This speaks to the affective and interpersonal implications of setting and communicating meeting purpose, which we explore in more detail in the following section. 

\subsubsection{Affective and Interpersonal Implications}\label{affectiveimplications}

Meetings with goals, or a lack of them, can cause people to experience feelings of (un)certainty, disrespect, and pressure; we address these respectively.

\paragraph{Certainty and Paranoia}

If the expected goals for a meeting are communicated, it allows attendees to predict what the meeting will be about. Five participants \pid{P2, P7, P8, P20, P21} described communicating goals as leading to clearer expectations.

Alternatively, a lack of clear goals can make a one-off meeting unpredictable. P18 and P15 both described feelings of uncertainty when invited to meetings without a clear goal or context. For P18, this was a concern about not being able to reply comprehensively to unexpected questions:

\begin{quote}
\iquote{``I won’t walk into a meeting where the attendee list is a combination I'm not expecting and if the meeting subject isn't really clear...I don't wanna come in and have the organizer say, OK, P18, give us the status on blah. Like whoa, I don't know what you're talking about.''} \pid{P18 - Customer Support, IC}
\end{quote}

For P15, a lack of clear goals for the meeting could lead to a feeling of paranoia. When brought into an ongoing meeting, where she had been told \iquote{``the general topic, but there was there was no goals''}, she described a sense of uncertainty and trepidation around her involvement in the meeting.

\begin{quote}
\iquote{``I also wasn't sure if I was being brought into this because they genuinely needed to know my perspective. Or was it? Were they trying to like, catch me in something?''} \pid{P15 - Customer Support, Manager}
\end{quote}

\paragraph{(Dis)respecting time}

When a meeting is understood to have a purpose, the organizer often uses this goal to justify taking up people’s time. They perceive people's time to be precious and limited, and so demonstrate respect by clearly justifying why the time is required for their meeting. 
\begin{quote}
  \iquote{ ``Being respectful of people's time, I'm always trying to think about what are actionable goals of this meeting.''} \pid{P3 - Administration, Manager}
\end{quote}

When asked if his attitude as an organizer differed from that of others, P18 responded: 
\begin{quote}
\iquote{``I'm more respectful of their schedule. To me, time is very important and if I'm gonna ask for somebody's time, I wanna have a reason for it.''} \pid{P18 - Customer Support, IC} 
\end{quote}

If a meeting lacks a goal outside of itself, some people perceive the meeting to be wasting their time.  
\begin{quote}
    \iquote{``I still really want to have topics beforehand because I often find that recurring meetings just turned into a waste of time because there's no set agenda.''} \pid{P15 - Customer Support, Manager} 
\end{quote}

After attending a meeting where there was \iquote{``no reason''} for them to be there, P12  \pid{Product Development, IC} reflected: \iquote{``I feel like my time is wasted, and I get a little salty about it.''}

Some people take it upon themselves to drive good meeting practice, if they perceive their time being wasted. However, taking this responsibility can result in feelings of discomfort \pid{P16, P21} or further resentment. Others mentioned that maintaining attention and engaging was more difficult in meetings that had no end-goal or clear agenda \pid{P7, P14, P17}.

\begin{quote} 

\iquote{``I'll kind of try to push them to practice good meeting habits, right? [...] It's tiring. Yeah, I'm just like I shouldn't have to do that.''} \pid{P16 - Product Development, Manager} 
\end{quote}
\begin{quote} 
\iquote{``I zone out. I don't listen, I'm chatting with other things, working on other things, multitasking, getting things done.''} \pid{P17 - Technical and Facilities, Manager}
\end{quote}

\paragraph{Pressure and Accountability}

If a meeting is seen to as a means to an end, there could be an expectation that attendees will commit to and act to achieve those ends. As a result, meetings with goals and expected outcomes can be seen as being more \iquote{``pressurizing''} \pid{P2 - Research, IC}.

P20 \pid{Administration, Manager} said she distributed timed agendas ahead of a meeting to ensure that individuals feel under pressure to keep to their own time: \iquote{``That's the guilt. That's the pressure that people have in their head.''}

While others perceive that setting and communicating goals can ensure accountability, they also acknowledged that this can make others uncomfortable or even \iquote{``fearful… because they want the freedom to just not be accountable''} \pid{P21 - Administration, Manager}. If some people want commitment to meeting outcomes, and others want to avoid committing, this can lead to understandable frustration on both sides \pid{P15, P21}.

\section{Discussion: Designing for Intentional Meetings}
\label{discussion}
As our findings have shown, people have a rich variety of intentions for their meetings, though the clarity and concreteness of these intentions vary across people and meetings, and some intentions may be dissonant. Current calendar and meeting software does not support people in reflecting on their intentions, nor does it include space designed to enable people to express their meeting intentions to others (See \autoref{fig:teaser}). For many in this organization (and, we surmise, many other organizations), time pressures are extreme, and so meetings have become a too-convenient container with the façade of easy scheduling and dynamism in the moment \cite{bergmann_meeting_2023}. As such, the status quo is to \textit{have meetings}, but then leaves people to manage their intentions for those meetings either implicitly or via ad-hoc methods which differ across people, teams, and organizations, and are disconnected from calendar and meeting systems. As a result, intentions for a meeting, expressed as `meeting goals’, are often communicated poorly, not at all, or misunderstood \cite{lopez-fresno_what_2022,geimer_meetings_2015}. The problem is exacerbated by the fragmentation of information workers’ attention across multiple tasks and software and the demands on them to manage and respond to queries across these spaces. This fragmentation means that \textit{intentional} behavior is de-prioritized in favor of \textit{habitual} behavior: responding to emails, meetings, calendar invites, chat messages, and task trackers. This leads to a counterproductive cycle of less intentionality and more reliance on default ways of working. 

We argue that software should support workers in regularly \textit{reflecting} on their intentions for meetings, acting on them, and where relevant, \textit{communicating} their intentions to others. Designing for intentionality is about creating tools for thought, as imagined by Licklider, Engelbart, and others \cite{rheingold_tools_1985}. \citet{norman_things_1993} argues that designing technology to encourage more `reflective cognition’ can increase intentional decision-making. %
In a way, this accords with “slow technology” that encourages people to slow down \cite{hallnas_slow_2001,baumer_reflective_2015}; however, rather than simply contrasting it with “fast” technology that promotes efficiency, %
slowing down can also promote efficiency by providing the time to make more intentional choices \cite{reicherts_make_2020,reicherts_extending_2022}. 

Supporting intentionality does not mean that every meeting should have an explicitly communicated set of goals. For some meetings, it may be best that goals are implicitly understood. However, supporting intentionality encourages people to reflect on meetings and meeting cultures, as organizers or attendees, to ensure that intentions are followed through in whatever way is appropriate. 

Once overarching intentions are surfaced and prioritized, they can help people select the appropriate tools, rather than reach for defaults. One version of this is making intentional choices about when meetings or asynchronous modalities may be better suited to goals. Asynchronous modes may give people more time to consider details, relevant connected issues, and alternatives, and they may ultimately be more time-efficient because composing a message and reading a message take very different amounts of time. They are also more searchable in the moment and over time, reducing the burden of some spontaneous meetings. A team that has a place that everyone is expected to update regularly and is considered the ground truth should be able to reserve meetings for time-critical issues, reducing the burden of meetings both for those who fume at goalless get-togethers and for those who find meetings difficult due to personal preference or differing abilities. 

The best meetings promote activities that benefit from the fast give-and-take and emergent content of interactivity (especially when remote or hybrid). Indeed, sometimes issues arising in an extended email thread may be best resolved through a meeting. Similarly, many workers value meetings for building relationships. However, the cadence and location of meetings are also a relevant part of this choice. Companies that wish to succeed at fully remote or hybrid work might find that maintaining relationships (especially internal relationships) may be better achieved via in-person events held at intervals of months to quarters (logistics permitting), than weekly or fortnightly online meetings held only for that purpose even while poorly suited to it~\cite{bergmann_meeting_2023}. Relatedly, many of those who resist the concept of communicating specific goals for meetings intended to be about open and emergent sharing may be setting themselves up for difficulties if attendees find that a \textit{place} to talk is not a \textit{reason} to talk. 

\subsection{When, where, and how to support intentionality?}

\begin{figure*}[h]
  \centering
  \includegraphics[width=\textwidth]{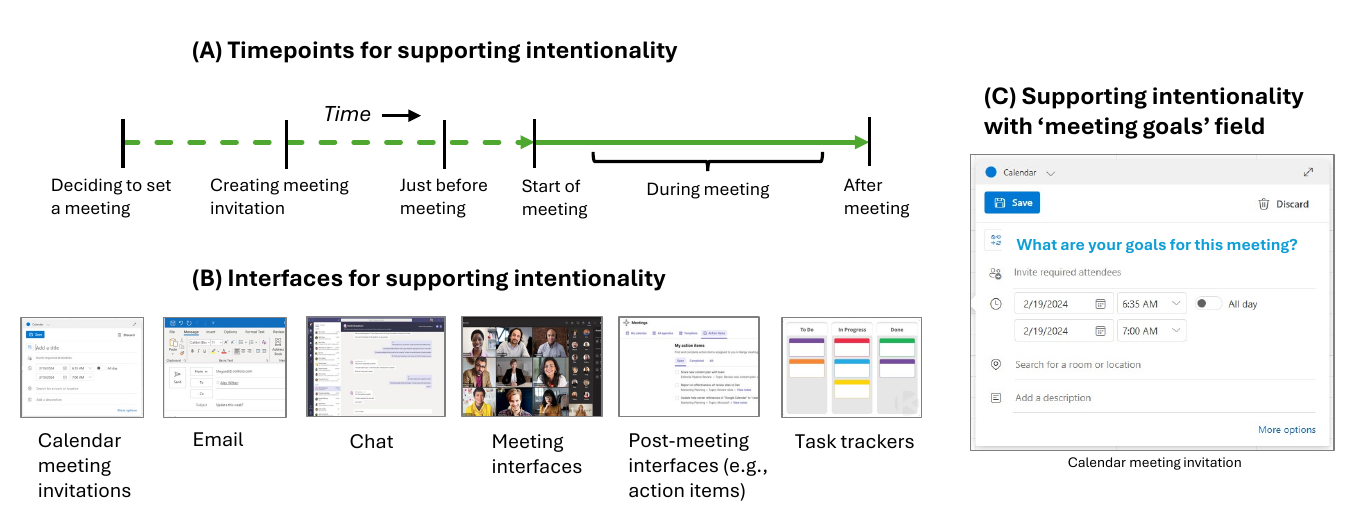}
  \caption{Designing for intentionality. There are many timepoints (A) and interfaces (B) where intentionality can be probed and surfaced. An example is a `meeting goals' field in a calendar invitation (C), used as an exploratory probe during our interviews.}
  \Description{The figure is divided into three main parts, labeled A, B, and C, detailing how to support intentionality in meetings using different interfaces at various time points. Part A, labeled 'Timepoints for supporting intentionality', is a horizontal timeline with dashed lines indicating the time before a meeting and a solid line for during a meeting. The timeline is marked with specific points: 'Deciding to set a meeting', 'Creating meeting invitation', 'Just before meeting', 'Start of meeting', 'During meeting', and 'After meeting'. Part B, labeled 'Interfaces for supporting intentionality', shows a series of interface screenshots. These include 'Calendar meeting invitations', 'Email', 'Chat', 'Meeting interfaces', 'Post-meeting interfaces (e.g., action items)', and 'Task trackers'. Part C, labeled 'Supporting intentionality with ‘meeting goals’ field', is a screenshot of a 'Calendar meeting invitation' interface. A dialog box titled 'What are your goals for this meeting?' is shown with fields to invite required attendees, set date and time, and additional options like 'Search for a room or location' and 'Add a description'. There are buttons to 'Save' or 'Discard' the invitation. The figure illustrates the use of various digital interfaces to organize and conduct meetings effectively, emphasizing the importance of setting goals throughout the meeting lifecycle.}
  \label{fig:intentionality}
\end{figure*}

As we have highlighted throughout the analysis, there are multiple time points and interfaces where the system could support intentionality, either by encouraging self-reflection to clarify intentions, explicit communication of intentions, or both. Temporally, people’s intentions can be probed when people first decide to set a meeting, at the time of meeting invitation creation, one or two days before a meeting, just prior to the meeting start time, at the start of a meeting, throughout the meeting, and even after the meeting ends (\autoref{fig:intentionality}A). If appropriate, these intentions can be translated into explicit goals that can be communicated to others via interfaces including calendar invitations, emails, chat messages, task managers, and the meeting interface itself (\autoref{fig:intentionality}B).  

\subsubsection{Value of a goals field in meeting scheduling}\label{meetinggoalsfield}
Surfacing intentionality can have different impacts at different time points. From the responses above, the most critical time is when an organizer is thinking about how to achieve work and has decided to schedule a meeting. Calendar invitation interfaces have numerous potential entry points for setting meetings, and so an intervention at the time of scheduling could help people decide on whether a meeting is the right tool to achieve their goals, and if so, communicate the reasons for it to attendees. As we mentioned in our introduction, as an exploratory probe in our interviews, we asked participants to consider a `meeting goals’ field in the context of calendar and meeting software (e.g., \autoref{fig:intentionality}C). We asked them to imagine the idea and respond and then asked them if they would prefer free text or one of the sets of options for meeting goals that we used as part of the initial interview quantitative probe (see Supplementary Materials for details). Here we outline their responses to round out the value of this simple feature.

Overall, participants thought this would be helpful for meeting organizers \pid{P18, P20}, and attendees \pid{P6, P14}. If a meeting is indeed appropriate, it can help people consider an agenda and other pre-meeting material to support their meeting goals. A meeting goals field may ensure that organizers have a \iquote{``very specific purpose in mind''} \pid{P20 - Administration, Manager}, and \iquote{``clarify why they’re doing it''} \pid{P18 - Customer Support, IC}, which may be particularly valuable when people view meeting goals as a set of discussion topics (as per Section \ref{subsubsec:meetingsasend}). In short, it can help people distinguish between the \textit{what} and \textit{why} of their meetings, and communicate this to others. This could help in \iquote{``aligning expectations''} between organizers and attendees, and help attendees know \iquote{``what to expect from their side''} \pid{P20 - Administration, Manager}, thereby providing certainty to attendees and preventing potential paranoia driven by a lack of information (as per §\ref{affectiveimplications}). An explicit meeting goals field could allow participants to review meeting goals and agendas before they accept \pid{P14 - Product Development, IC}, and offer a consistent place for attendees to \iquote{``always look at this meeting goal field for any meeting''} \pid{P6 - Product Development, Manager}. Thus, a meeting goals field can help (a) organizers to decide on whether they need a meeting in the first place, (b) invitees to decide on whether they should attend, and (c) attendees to adequately prepare for the meeting as needed (as per §\ref{functionalimplications}).

For organizers, some participants valued the idea of a meeting goals field due to its ability to prompt themselves amidst the fragmentation and time pressure of work \pid{P13, P6, P4}, a key obstacle to setting meeting goals identified in §\ref{obstacles}. P5 even felt that it should be mandatory. 

\begin{quote} 
\iquote{``People are not naturally thinking about meeting goals. [It's good to] to prompt them before they sense hey, by the way, you didn't send any meeting goals you know - would you like to check one of these?''} \pid{P6 - Product Development, Manager} 
\end{quote}
\begin{quote} 
\iquote{``…like you can't set up a meeting… if you don't set an agenda and outcomes.''} \pid{P5 - Product Development, IC} 
\end{quote}

For invitees, the base level value of an organizer filling out this field was proposed as understanding the broader aims \pid{P16}.  Beyond consuming the context, other participants felt that if an organizer had provided the goals, then invitees could actively collaborate on goal-setting \pid{P4}. This could also enable understanding how the work in the meeting will be apportioned \pid{P4}. Thus, there is an opportunity to lift the organizational burden from meeting organizers, particularly for large meetings with many different attendee roles (as found in §\ref{obstacles}).

\begin{quote}
\iquote{``…sharing the goals with the invite makes a lot of sense because it's good context.''} \pid{P16 - Product Development, Manager}
\end{quote}
\begin{quote}
\iquote{``…you get a notification in your chat that says <<Name>>'s outlined some goals for this meeting. Do you wanna review it and add your own?''} \pid{P4 - Sales, IC}
\end{quote}
\begin{quote}
\iquote{``It would have would be nice to know does my colleague plan on doing anything in the call or am I responsible for the full 30 minutes?''} \pid{P4 - Sales, IC}
\end{quote}

Just before or at the start of meetings, a filled meeting goals field can help people reaffirm goals \pid{P16, P13}, ensure clarity across attendees, and motivate the agenda as an enabler of the meetings’ goals. In this way, it could help meetings stay on track (as per §\ref{functionalimplications}).   

\begin{quote}
\iquote{``Usually the the meeting invitation is a little bit divorced from the meeting itself because I'm usually not sending it out immediately before. So I think I'm, because it would be there, I assume, when I reopened the meeting to start it yes, seeing my refresher of my goal is good.''} \pid{P13 - Research, IC} 
\end{quote}

This instantaneous reflection before a meeting could be particularly useful before attending an instance of a recurring meeting. In \autoref{fig:intentionality}B, the time between the creation of the invitation and the start of the meeting increases as the recurring meeting continues to run. As this time lengthens, the initial intention for the recurring meeting could be forgotten or become irrelevant. It could therefore be useful to reflect on the intentions for recurring meetings at regular intervals: 

\begin{quote}
\iquote{``If it was periodic, for instance, once a quarter, or once every half year or something, on a recurring meeting, then OK. Then then maybe there could be some value there to the meeting organizer.''} \pid{P7 - Product Development, IC}
\end{quote}

Prior meeting science research has pointed out that meeting training is often lacking in organizations~\cite{rogelberg2018surprising}. Several of our participants perceived that a lack of training in meeting practices acted as an obstacle to setting meeting goals \pid{P12, P17}. Embedding meeting intentionality into calendar and meeting interfaces could act to teach and/or reinforce these practices in people’s minds, so they clarify meeting purpose more automatically: 
\begin{quote}
\iquote{``It's kind of training the user, inadvertently, on how to run a meeting and how to make a meeting successful right?''} 
\pid{P12 - Product Development, IC}. 
\end{quote}

However, some participants were skeptical of a meeting goals field because it may be too pressurizing \pid{P2 - Research, IC}, neglected \pid{P1 - Technical and Facilities, IC}, or difficult to update \pid{P16 - Product Development, Manager}. Thus, although a meeting goals field has the potential to strengthen certain norms and counter differences in personality traits (as per §\ref{obstacles}), it may also succumb to these differences.    

\begin{quote}
\iquote{``I feel like people would just write, you know, garbage in there because for that meeting where it's not important and then at some point it's just gonna be like a field people just ignore.''} \pid{P1 - Technical and Facilities, IC}
\end{quote}
\begin{quote}
\iquote{``Once the meeting is shared, if those goals need to change or those goals need to be adjusted, it becomes harder for me to go in and edit them, right?''} \pid{P16 - Product Development, Manager} 
\end{quote}

If a meeting goal field is included and has a prospective value at the start of a meeting, it has a natural counterpart in retrospective value at the end of a meeting. After meetings, re-surfacing the proposed meeting goals can help people evaluate the meeting’s effectiveness, and broader project planning, including whether a follow-up meeting is necessary. It can also tie the meeting to future work and ensure clarity and agreement across collaborators: 

\begin{quote}\iquote{``It's like, hey, what are your goals for this meeting? Type, type, type, and then 5 minutes after and `How did you meet them all' or, we thought it met three of five.''} \pid{P16 - Product Development, Manager}.\end{quote}

Indeed, this speaks to the accountability afforded by setting clear meeting goals, and although some view this accountability as `pressurizing' (§\ref{implications}), a meeting goals field could help anticipate potential misunderstandings before meetings begin.  

\subsubsection{Using Generative AI to Promote Meeting Goals and Intentionality}

Generative AI offers an opportunity to support intentionality in an interactive and contextually aware manner (e.g., \cite{li_exploring_2023}). In the near future, generative AI can serve as a bridge across currently fragmented software and workflows by eliciting and reinforcing intentionality across workflow and project contexts---and old idea now made more plausible with generative AI to cope with the huge volume and variation of relevant materials in organizations~\cite{craven_goalsandprocesses_1995}. For example, LLMs can nudge people to self-reflect on their intentions for a meeting in the context of available information, such as their and others’ schedules, awareness of work, and project progress. The interactive chat style of an LLM may overcome the cold start problem of a blank field or a field that only suggests `write your meeting goal here'. Enriched with contextual data, LLMs have the potential to make currently ineffective dashboards (and the data therein) \cite{alhamadi_challenges_2020} come alive via conversation. 

Even without LLMs, self-reflection interventions in the workplace have shown promise in helping people reflect on and achieve their goals \cite{meyer_enabling_2021,kocielnik2018designing}. Other work has explored the design space of self-reflection, revealing a rich toolkit, including temporal information, conversations, and comparisons \cite{bentvelzen_revisiting_2022, baumer_reviewing_2014,baumer_reflective_2015}. Indeed, as evidenced by participants in §\ref{meetinggoalsfield}, a simple meeting goals field has the potential to help prompt or people to think more intentionally about their collaboration amidst the time pressure and routinization of work.   

Once intentions are clarified, generative AI could reinforce them in a contextually-aware manner across software and workflows. For example, based on people’s scheduling input, metadata, and invitee lists, LLMs could automate the creation of meeting goals and agendas, as suggested by some participants \pid{P9, P10}, and support people in following through on their intentions in meetings. This could help address the observed fragmentation and diffusion of meeting-related information across software and workflows and support users who face time constraints and who may differ in their personalities and work priorities (as per §\ref{obstacles} and §\ref{howgoalsarise}).  

\begin{quote}
\iquote{``Oh yeah! If you could just have some sort of AI automation when it comes to goals, that would probably save us a ton of time in my team''} \pid{P9 - Product Development, IC}
\end{quote}
\begin{quote}
\iquote{``I think that if it could help, like based on if I have a really good name of a meeting and maybe I have like an agenda, like if it was smart enough to say the goal should be this and make recommendations that could change, I would use that, yeah.''} \pid{P10 - Sales, IC} 
\end{quote}

This support could be in a manner akin to current notions of private or meeting-wide facilitator bots (e.g. \cite{kim_meetingbot_2020}), but generative AI holds the promise of far more variety in methods for helping people consider goals.

For example, at the point of setting a meeting, generative AI might help an organizer visualize a set of alternative methods for achieving a goal, involving a meeting or not, or, more far-reaching, suggest different cadences of meetings or balances of synchronous and asynchronous work. Generative AI's special value here is in providing users with a range of contextually-informed ways of understanding their alternate paths, from simple lists to flowcharts and timelines, to cartoon or realistic storyboards, or even more abstract concepts. Such systems could also track goal achievement and suggest changes or even create and monitor experiments for teams to try achieving goals faster or with less stress. They may offer an opportunity for teams consisting of people with different personalities, expectations, and priorities (as per §\ref{obstacles}) to find agreeable and effective ways of collaborating. Moreover, generative AI could be used to offer clarifications on meeting goals. Attendees could pose questions about the meeting, confident that they will not inadvertently offend the organizer. This could foster an environment of openness and information-sharing, tackling the obstacle of psychological safety in clarifying meeting goals (as per §\ref{obstacles}).

In the more distant future, generative AI offers an opportunity to re-imagine meeting interfaces entirely to be goal-driven rather than AV-driven, reducing the one-size-fits-all meeting interfaces that have led in part to videoconferencing fatigue. Interfaces that are responsive to people’s intentions and relevant contexts could be generated on the fly, such that an update round, a brainstorm, a sales pitch, a brown-bag lunch, a happy hour, a games night, etc. could all feature different configurations and representations of people and resources, while also taking into account team and organizational culture, and even ethnic and sub-group cultural norms and preferences. Such interfaces could also change to fit the different internal phases of an activity, e.g. in an educational context, a class period could transition people and resource configurations from lecture to breakout to debate and summary phases. In any generated interface, and across phases, then, goal-directed behavior would drive the look and feel of the experience, making goal-directed behavior explicit in meetings themselves and making explicit where any given meeting fits into larger project or team goals. This would not only help meetings stay on track and therefore be more effective, but, due to generative AI's flexibility, can also enable dynamically balancing \textit{adaptability and discovery} with \textit{explicit goal setting} in meetings (as per §\ref{functionalimplications}).   

While our work investigates meeting goals across a wide range of work areas, meeting intentionality is likely to be highly specific to the particular work being done. For example, customer support and sales teams have to adapt to the practices of external customers and companies, and their workflow is often directed by explicit quotas and targets in a way that research is not. Future foundational research can inform the design of generative AI systems so they can deftly and intelligently support these varying work practices.

\subsubsection{Challenges of Intentionality Interventions}
Embedding intentionality into meeting and collaboration systems may create new demands on users' time and attention, and increase the pressure they feel at work.

In time-pressured work environments, users may perceive a system's cues for active intentionality to be another drain on their cognitive resources and time. As scheduling meetings requires extensive logistical effort \cite{sun_rhythm_2023}, the key challenge is that cues for intentionality may be simply ignored by organizers. In the days prior to a meeting, both organizers and attendees in time-pressured work environments are likely to prioritize other acute work needs over reflection on a future meeting's goals. And, again, in the moments immediately before a meeting or at the meeting starts, the rush to just start the meeting may well overpower the perceived need to reflect on its purpose. Dealing with this challenge—an instance of the psychological phenomenon of temporal discounting ~\cite{bulley_deliberating_2020}—thus requires design that actively provides a sense of balancing immediate time taken against later time saved.

During a meeting, an over-analysis of intention and goals could distract from the content of the meeting, resulting in fewer decisions being made and an over-critical stance on how time is being used. The risk of `paralysis-by-analysis' has been described in research into management ~\cite{langley1995between} and planning systems~\cite{lenz1985paralysis}. Rumination is often highlighted as a risk of reflective practices, if one cannot find a solution to a problem, but continues to think about it~\cite{eikey2021beyond, trapnell1999private}. As well as reminding users to be intentional, systems should guide them to construct and articulate their intentions in a productive way, so as to avoid negative cycles of introspection and analysis. This challenge also has clear links to the division of attention that is already occurring in online meetings—for example, the need to attend to both the Audio/Video of the meeting and the meeting chat~\cite{sarkar_promise_2021}. Dealing with this challenge thus requires design that dynamically adjusts peripheral and focused attention on goal needs.

By repeatedly surfacing intentionality, systems could be perceived as applying pressure on users, contributing to greater feelings of stress at work~\cite{mehrab-czerwinski_stressatwork_2022}. In the context of meetings, demanding an explicit statement of intentionality could undermine a purpose of `connecting and being there', resulting in a reduction in the number of meetings with this purpose, or causing them to feel transactional or forced. Further, in the worst case, promoting a culture of intentionality as accounted for via technology might lead to employees to feel that they are under workplace surveillance~\cite{mcparland_dataveillance_2020}. Dealing with these challenges will require at least as much workplace policy and cultural management as technological design solutions to engender trust, because they get to the heart of what it means to work in a particular organization. 

\section{Conclusion}
\label{conclusion}

Existing calendar and meeting systems provide canvases for planning and connection, but have overlooked the crucial need to assist workers in clarifying the purpose of meetings. This is an issue because workers may organize or attend meetings based on two quite different mental models: meetings as a means to an end, and meetings as an end in themselves. We thus conceptualize meetings as mediating intentionality or as directly representing an intention. Intentionality travels through the meeting lifecycle, from its conceptualization to its execution. However, this intentionality can be made material or obstructed at different moments. Obstructions, clearly, may have negative impacts, including anxiety, resentment, and time wastage. This may be recognized by organizers and attendees, and while some workers are excellent at making their intentionality material and bringing others along with them, many others feel trapped in a status quo of too many meetings that lack a clear purpose. 

We argue that the fragmented ecosystem of calendaring, planning, and meeting technologies needs to do more to catalyze intentionality. Intentionality should be made explicit across the meeting lifecycle, such that goal-directed behavior is habituated. Intentionality should also be used to drive the design of new features in planning, calendar, and meeting interfaces that augment intentionality as an accountable part of work, and may even be manifested in interfaces that adapt and customize themselves to intentionality.

\begin{acks}
We thank all the participants for sharing their time and perspectives with us, and the reviewers for their constructive feedback. 
\end{acks}

\bibliographystyle{ACM-Reference-Format}
\bibliography{CHI2024_Goals}


\begin{thebibliography}{86}


\ifx \showCODEN    \undefined \def \showCODEN     #1{\unskip}     \fi
\ifx \showDOI      \undefined \def \showDOI       #1{#1}\fi
\ifx \showISBNx    \undefined \def \showISBNx     #1{\unskip}     \fi
\ifx \showISBNxiii \undefined \def \showISBNxiii  #1{\unskip}     \fi
\ifx \showISSN     \undefined \def \showISSN      #1{\unskip}     \fi
\ifx \showLCCN     \undefined \def \showLCCN      #1{\unskip}     \fi
\ifx \shownote     \undefined \def \shownote      #1{#1}          \fi
\ifx \showarticletitle \undefined \def \showarticletitle #1{#1}   \fi
\ifx \showURL      \undefined \def \showURL       {\relax}        \fi
\providecommand\bibfield[2]{#2}
\providecommand\bibinfo[2]{#2}
\providecommand\natexlab[1]{#1}
\providecommand\showeprint[2][]{arXiv:#2}

\bibitem[Alhamadi(2020)]%
        {alhamadi_challenges_2020}
\bibfield{author}{\bibinfo{person}{Mohammed Alhamadi}.} \bibinfo{year}{2020}\natexlab{}.
\newblock \showarticletitle{Challenges, {Strategies} and {Adaptations} on {Interactive} {Dashboards}}. In \bibinfo{booktitle}{\emph{Proceedings of the 28th {ACM} {Conference} on {User} {Modeling}, {Adaptation} and {Personalization}}} \emph{(\bibinfo{series}{{UMAP} '20})}. \bibinfo{publisher}{Association for Computing Machinery}, \bibinfo{address}{New York, NY, USA}, \bibinfo{pages}{368--371}.
\newblock
\showISBNx{978-1-4503-6861-2}
\urldef\tempurl%
\url{https://doi.org/10.1145/3340631.3398678}
\showDOI{\tempurl}


\bibitem[Allen et~al\mbox{.}(2014)]%
        {allen_understanding_2014}
\bibfield{author}{\bibinfo{person}{Joseph~A. Allen}, \bibinfo{person}{Tammy Beck}, \bibinfo{person}{Cliff W.~Scott}, {and} \bibinfo{person}{Steven G.~Rogelberg}.} \bibinfo{year}{2014}\natexlab{}.
\newblock \showarticletitle{Understanding workplace meetings: {A} qualitative taxonomy of meeting purposes}.
\newblock \bibinfo{journal}{\emph{Management Research Review}} \bibinfo{volume}{37}, \bibinfo{number}{9} (\bibinfo{date}{Jan.} \bibinfo{year}{2014}), \bibinfo{pages}{791--814}.
\newblock
\showISSN{2040-8269}
\urldef\tempurl%
\url{https://doi.org/10.1108/MRR-03-2013-0067}
\showDOI{\tempurl}
\newblock
\shownote{Publisher: Emerald Group Publishing Limited}.


\bibitem[Bang et~al\mbox{.}(2010)]%
        {bang_effectiveness_2010}
\bibfield{author}{\bibinfo{person}{Henning Bang}, \bibinfo{person}{Synne~L. Fuglesang}, \bibinfo{person}{Mariann~R. Ovesen}, {and} \bibinfo{person}{Dag~Erik Eilertsen}.} \bibinfo{year}{2010}\natexlab{}.
\newblock \showarticletitle{Effectiveness in top management group meetings: {The} role of goal clarity, focused communication, and learning behavior: {Effectiveness} in top management meetings}.
\newblock \bibinfo{journal}{\emph{Scandinavian Journal of Psychology}} \bibinfo{volume}{51}, \bibinfo{number}{3} (\bibinfo{date}{Jan.} \bibinfo{year}{2010}), \bibinfo{pages}{253--261}.
\newblock
\showISSN{00365564, 14679450}
\urldef\tempurl%
\url{https://doi.org/10.1111/j.1467-9450.2009.00769.x}
\showDOI{\tempurl}


\bibitem[Baumer(2015)]%
        {baumer_reflective_2015}
\bibfield{author}{\bibinfo{person}{Eric~P.S. Baumer}.} \bibinfo{year}{2015}\natexlab{}.
\newblock \showarticletitle{Reflective {Informatics}: {Conceptual} {Dimensions} for {Designing} {Technologies} of {Reflection}}. In \bibinfo{booktitle}{\emph{Proceedings of the 33rd {Annual} {ACM} {Conference} on {Human} {Factors} in {Computing} {Systems}}} \emph{(\bibinfo{series}{{CHI} '15})}. \bibinfo{publisher}{Association for Computing Machinery}, \bibinfo{address}{New York, NY, USA}, \bibinfo{pages}{585--594}.
\newblock
\showISBNx{978-1-4503-3145-6}
\urldef\tempurl%
\url{https://doi.org/10.1145/2702123.2702234}
\showDOI{\tempurl}


\bibitem[Baumer et~al\mbox{.}(2014)]%
        {baumer_reviewing_2014}
\bibfield{author}{\bibinfo{person}{Eric~P.S. Baumer}, \bibinfo{person}{Vera Khovanskaya}, \bibinfo{person}{Mark Matthews}, \bibinfo{person}{Lindsay Reynolds}, \bibinfo{person}{Victoria Schwanda~Sosik}, {and} \bibinfo{person}{Geri Gay}.} \bibinfo{year}{2014}\natexlab{}.
\newblock \showarticletitle{Reviewing reflection: on the use of reflection in interactive system design}. In \bibinfo{booktitle}{\emph{Proceedings of the 2014 conference on {Designing} interactive systems}} \emph{(\bibinfo{series}{{DIS} '14})}. \bibinfo{publisher}{Association for Computing Machinery}, \bibinfo{address}{New York, NY, USA}, \bibinfo{pages}{93--102}.
\newblock
\showISBNx{978-1-4503-2902-6}
\urldef\tempurl%
\url{https://doi.org/10.1145/2598510.2598598}
\showDOI{\tempurl}


\bibitem[Bentvelzen et~al\mbox{.}(2022)]%
        {bentvelzen_revisiting_2022}
\bibfield{author}{\bibinfo{person}{Marit Bentvelzen}, \bibinfo{person}{Paweł~W. Woźniak}, \bibinfo{person}{Pia~S.F. Herbes}, \bibinfo{person}{Evropi Stefanidi}, {and} \bibinfo{person}{Jasmin Niess}.} \bibinfo{year}{2022}\natexlab{}.
\newblock \showarticletitle{Revisiting {Reflection} in {HCI}: {Four} {Design} {Resources} for {Technologies} that {Support} {Reflection}}.
\newblock \bibinfo{journal}{\emph{Proceedings of the ACM on Interactive, Mobile, Wearable and Ubiquitous Technologies}} \bibinfo{volume}{6}, \bibinfo{number}{1} (\bibinfo{date}{March} \bibinfo{year}{2022}), \bibinfo{pages}{2:1--2:27}.
\newblock
\urldef\tempurl%
\url{https://doi.org/10.1145/3517233}
\showDOI{\tempurl}


\bibitem[Bergmann et~al\mbox{.}(2023)]%
        {bergmann_meeting_2023}
\bibfield{author}{\bibinfo{person}{Rachel Bergmann}, \bibinfo{person}{Sean Rintel}, \bibinfo{person}{Nancy Baym}, \bibinfo{person}{Advait Sarkar}, \bibinfo{person}{Damian Borowiec}, \bibinfo{person}{Priscilla Wong}, {and} \bibinfo{person}{Abigail Sellen}.} \bibinfo{year}{2023}\natexlab{}.
\newblock \showarticletitle{Meeting (the) {Pandemic}: {Videoconferencing} {Fatigue} and {Evolving} {Tensions} of {Sociality} in {Enterprise} {Video} {Meetings} {During} {COVID}-19}.
\newblock \bibinfo{journal}{\emph{Computer Supported Cooperative Work (CSCW)}} \bibinfo{volume}{32}, \bibinfo{number}{2} (\bibinfo{date}{June} \bibinfo{year}{2023}), \bibinfo{pages}{347--383}.
\newblock
\showISSN{0925-9724, 1573-7551}
\urldef\tempurl%
\url{https://doi.org/10.1007/s10606-022-09451-6}
\showDOI{\tempurl}


\bibitem[Braun and Clarke(2006)]%
        {braun2006using}
\bibfield{author}{\bibinfo{person}{Virginia Braun} {and} \bibinfo{person}{Victoria Clarke}.} \bibinfo{year}{2006}\natexlab{}.
\newblock \showarticletitle{Using thematic analysis in psychology}.
\newblock \bibinfo{journal}{\emph{Qualitative research in psychology}} \bibinfo{volume}{3}, \bibinfo{number}{2} (\bibinfo{year}{2006}), \bibinfo{pages}{77--101}.
\newblock


\bibitem[Bulley and Schacter(2020)]%
        {bulley_deliberating_2020}
\bibfield{author}{\bibinfo{person}{Adam Bulley} {and} \bibinfo{person}{Daniel~L. Schacter}.} \bibinfo{year}{2020}\natexlab{}.
\newblock \showarticletitle{Deliberating trade-offs with the future}.
\newblock \bibinfo{journal}{\emph{Nature human behaviour}} \bibinfo{volume}{4}, \bibinfo{number}{3} (\bibinfo{date}{March} \bibinfo{year}{2020}), \bibinfo{pages}{238--247}.
\newblock
\showISSN{2397-3374}
\urldef\tempurl%
\url{https://doi.org/10.1038/s41562-020-0834-9}
\showDOI{\tempurl}


\bibitem[Cao et~al\mbox{.}(2021)]%
        {cao2021large}
\bibfield{author}{\bibinfo{person}{Hancheng Cao}, \bibinfo{person}{Chia-Jung Lee}, \bibinfo{person}{Shamsi Iqbal}, \bibinfo{person}{Mary Czerwinski}, \bibinfo{person}{Priscilla~NY Wong}, \bibinfo{person}{Sean Rintel}, \bibinfo{person}{Brent Hecht}, \bibinfo{person}{Jaime Teevan}, {and} \bibinfo{person}{Longqi Yang}.} \bibinfo{year}{2021}\natexlab{}.
\newblock \showarticletitle{Large scale analysis of multitasking behavior during remote meetings}. In \bibinfo{booktitle}{\emph{Proceedings of the 2021 CHI Conference on Human Factors in Computing Systems}}. \bibinfo{pages}{1--13}.
\newblock


\bibitem[Cohen et~al\mbox{.}(2011)]%
        {cohen_meeting_2011}
\bibfield{author}{\bibinfo{person}{Melissa~A. Cohen}, \bibinfo{person}{Steven~G. Rogelberg}, \bibinfo{person}{Joseph~A. Allen}, {and} \bibinfo{person}{Alexandra Luong}.} \bibinfo{year}{2011}\natexlab{}.
\newblock \showarticletitle{Meeting design characteristics and attendee perceptions of staff/team meeting quality}.
\newblock \bibinfo{journal}{\emph{Group Dynamics: Theory, Research, and Practice}} \bibinfo{volume}{15}, \bibinfo{number}{1} (\bibinfo{year}{2011}), \bibinfo{pages}{90--104}.
\newblock
\showISSN{1930-7802}
\urldef\tempurl%
\url{https://doi.org/10.1037/a0021549}
\showDOI{\tempurl}
\newblock
\shownote{Place: US Publisher: Educational Publishing Foundation}.


\bibitem[Craven and Mahling(1995)]%
        {craven_goalsandprocesses_1995}
\bibfield{author}{\bibinfo{person}{No\"{e}l Craven} {and} \bibinfo{person}{Dirk Mahling}.} \bibinfo{year}{1995}\natexlab{}.
\newblock \showarticletitle{Goals and Processes: A Task Basis for Projects and Workflows}. In \bibinfo{booktitle}{\emph{Proceedings of Conference on Organizational Computing Systems}} (Milpitas, California, USA) \emph{(\bibinfo{series}{COCS '95})}. \bibinfo{publisher}{Association for Computing Machinery}, \bibinfo{address}{New York, NY, USA}, \bibinfo{pages}{237–248}.
\newblock
\showISBNx{0897917065}
\urldef\tempurl%
\url{https://doi.org/10.1145/224019.224045}
\showDOI{\tempurl}


\bibitem[Cutler et~al\mbox{.}(2021)]%
        {cutler_meeting_2021}
\bibfield{author}{\bibinfo{person}{Ross Cutler}, \bibinfo{person}{Yasaman Hosseinkashi}, \bibinfo{person}{Jamie Pool}, \bibinfo{person}{Senja Filipi}, \bibinfo{person}{Robert Aichner}, \bibinfo{person}{Yuan Tu}, {and} \bibinfo{person}{Johannes Gehrke}.} \bibinfo{year}{2021}\natexlab{}.
\newblock \showarticletitle{Meeting {Effectiveness} and {Inclusiveness} in {Remote} {Collaboration}}.
\newblock \bibinfo{journal}{\emph{Proceedings of the ACM on Human-Computer Interaction}} \bibinfo{volume}{5}, \bibinfo{number}{CSCW1} (\bibinfo{date}{April} \bibinfo{year}{2021}), \bibinfo{pages}{1--29}.
\newblock
\showISSN{2573-0142}
\urldef\tempurl%
\url{https://doi.org/10.1145/3449247}
\showDOI{\tempurl}


\bibitem[De~Vreede et~al\mbox{.}(2003)]%
        {de_vreede_how_2003}
\bibfield{author}{\bibinfo{person}{Gert-Jan De~Vreede}, \bibinfo{person}{Robert~M. Davison}, {and} \bibinfo{person}{Robert~O. Briggs}.} \bibinfo{year}{2003}\natexlab{}.
\newblock \showarticletitle{How a silver bullet may lose its shine}.
\newblock \bibinfo{journal}{\emph{Commun. ACM}} \bibinfo{volume}{46}, \bibinfo{number}{8} (\bibinfo{date}{Aug.} \bibinfo{year}{2003}), \bibinfo{pages}{96--101}.
\newblock
\showISSN{0001-0782, 1557-7317}
\urldef\tempurl%
\url{https://doi.org/10.1145/859670.859676}
\showDOI{\tempurl}


\bibitem[Döring et~al\mbox{.}(2022)]%
        {doring_videoconference_2022}
\bibfield{author}{\bibinfo{person}{Nicola Döring}, \bibinfo{person}{Katrien~De Moor}, \bibinfo{person}{Markus Fiedler}, \bibinfo{person}{Katrin Schoenenberg}, {and} \bibinfo{person}{Alexander Raake}.} \bibinfo{year}{2022}\natexlab{}.
\newblock \showarticletitle{Videoconference {Fatigue}: {A} {Conceptual} {Analysis}}.
\newblock \bibinfo{journal}{\emph{International Journal of Environmental Research and Public Health}} \bibinfo{volume}{19}, \bibinfo{number}{4} (\bibinfo{date}{Feb.} \bibinfo{year}{2022}), \bibinfo{pages}{2061}.
\newblock
\showISSN{1660-4601}
\urldef\tempurl%
\url{https://doi.org/10.3390/ijerph19042061}
\showDOI{\tempurl}


\bibitem[Eikey et~al\mbox{.}(2021)]%
        {eikey2021beyond}
\bibfield{author}{\bibinfo{person}{Elizabeth~Victoria Eikey}, \bibinfo{person}{Clara~Marques Caldeira}, \bibinfo{person}{Mayara~Costa Figueiredo}, \bibinfo{person}{Yunan Chen}, \bibinfo{person}{Jessica~L Borelli}, \bibinfo{person}{Melissa Mazmanian}, {and} \bibinfo{person}{Kai Zheng}.} \bibinfo{year}{2021}\natexlab{}.
\newblock \showarticletitle{Beyond self-reflection: introducing the concept of rumination in personal informatics}.
\newblock \bibinfo{journal}{\emph{Personal and Ubiquitous Computing}} \bibinfo{volume}{25}, \bibinfo{number}{3} (\bibinfo{year}{2021}), \bibinfo{pages}{601--616}.
\newblock


\bibitem[Etikan et~al\mbox{.}(2016)]%
        {etikan2016comparison}
\bibfield{author}{\bibinfo{person}{Ilker Etikan}, \bibinfo{person}{Sulaiman~Abubakar Musa}, \bibinfo{person}{Rukayya~Sunusi Alkassim}, {et~al\mbox{.}}} \bibinfo{year}{2016}\natexlab{}.
\newblock \showarticletitle{Comparison of convenience sampling and purposive sampling}.
\newblock \bibinfo{journal}{\emph{American journal of theoretical and applied statistics}} \bibinfo{volume}{5}, \bibinfo{number}{1} (\bibinfo{year}{2016}), \bibinfo{pages}{1--4}.
\newblock


\bibitem[Finn et~al\mbox{.}(1997)]%
        {finn-sellen-wilbur_vmc_1997}
\bibfield{editor}{\bibinfo{person}{K Finn}, \bibinfo{person}{Abigail Sellen}, {and} \bibinfo{person}{S Wilbur}} (Eds.). \bibinfo{year}{1997}\natexlab{}.
\newblock \bibinfo{booktitle}{\emph{Video-mediated communication}}.
\newblock \bibinfo{publisher}{Erlbaum}, \bibinfo{address}{Mahwah, NJ}.
\newblock


\bibitem[Geimer et~al\mbox{.}(2015)]%
        {geimer_meetings_2015}
\bibfield{author}{\bibinfo{person}{Jennifer~L. Geimer}, \bibinfo{person}{Desmond~J. Leach}, \bibinfo{person}{Justin~A. DeSimone}, \bibinfo{person}{Steven~G. Rogelberg}, {and} \bibinfo{person}{Peter~B. Warr}.} \bibinfo{year}{2015}\natexlab{}.
\newblock \showarticletitle{Meetings at work: {Perceived} effectiveness and recommended improvements}.
\newblock \bibinfo{journal}{\emph{Journal of Business Research}} \bibinfo{volume}{68}, \bibinfo{number}{9} (\bibinfo{date}{Sept.} \bibinfo{year}{2015}), \bibinfo{pages}{2015--2026}.
\newblock
\showISSN{01482963}
\urldef\tempurl%
\url{https://doi.org/10.1016/j.jbusres.2015.02.015}
\showDOI{\tempurl}


\bibitem[Gollwitzer and Sheeran(2006)]%
        {gollwitzer_implementation_2006}
\bibfield{author}{\bibinfo{person}{Peter~M. Gollwitzer} {and} \bibinfo{person}{Paschal Sheeran}.} \bibinfo{year}{2006}\natexlab{}.
\newblock \showarticletitle{Implementation {Intentions} and {Goal} {Achievement}: {A} {Meta}‐analysis of {Effects} and {Processes}}.
\newblock In \bibinfo{booktitle}{\emph{Advances in {Experimental} {Social} {Psychology}}}. Vol.~\bibinfo{volume}{38}. \bibinfo{publisher}{Elsevier}, \bibinfo{pages}{69--119}.
\newblock
\showISBNx{978-0-12-015238-4}
\urldef\tempurl%
\url{https://doi.org/10.1016/S0065-2601(06)38002-1}
\showDOI{\tempurl}


\bibitem[Gonzalez~Diaz et~al\mbox{.}(2022)]%
        {gonzalez_diaz_making_2022}
\bibfield{author}{\bibinfo{person}{Carlos Gonzalez~Diaz}, \bibinfo{person}{John Tang}, \bibinfo{person}{Advait Sarkar}, {and} \bibinfo{person}{Sean Rintel}.} \bibinfo{year}{2022}\natexlab{}.
\newblock \showarticletitle{Making {Space} for {Social} {Time}: {Supporting} {Conversational} {Transitions} {Before}, {During}, and {After} {Video} {Meetings}}. In \bibinfo{booktitle}{\emph{2022 {Symposium} on {Human}-{Computer} {Interaction} for {Work}}} \emph{(\bibinfo{series}{{CHIWORK} 2022})}. \bibinfo{publisher}{Association for Computing Machinery}, \bibinfo{address}{New York, NY, USA}, \bibinfo{pages}{1--11}.
\newblock
\showISBNx{978-1-4503-9655-4}
\urldef\tempurl%
\url{https://doi.org/10.1145/3533406.3533417}
\showDOI{\tempurl}


\bibitem[Grudin(1994)]%
        {grudin_CSCWreview_1994}
\bibfield{author}{\bibinfo{person}{J. Grudin}.} \bibinfo{year}{1994}\natexlab{}.
\newblock \showarticletitle{Computer-supported cooperative work: history and focus}.
\newblock \bibinfo{journal}{\emph{Computer}} \bibinfo{volume}{27}, \bibinfo{number}{5} (\bibinfo{year}{1994}), \bibinfo{pages}{19--26}.
\newblock
\urldef\tempurl%
\url{https://doi.org/10.1109/2.291294}
\showDOI{\tempurl}


\bibitem[Hallnäs and Redström(2001)]%
        {hallnas_slow_2001}
\bibfield{author}{\bibinfo{person}{Lars Hallnäs} {and} \bibinfo{person}{Johan Redström}.} \bibinfo{year}{2001}\natexlab{}.
\newblock \showarticletitle{Slow {Technology} – {Designing} for {Reflection}}.
\newblock \bibinfo{journal}{\emph{Personal and Ubiquitous Computing}} \bibinfo{volume}{5}, \bibinfo{number}{3} (\bibinfo{date}{Aug.} \bibinfo{year}{2001}), \bibinfo{pages}{201--212}.
\newblock
\showISSN{1617-4909}
\urldef\tempurl%
\url{https://doi.org/10.1007/PL00000019}
\showDOI{\tempurl}


\bibitem[Harrison(2009)]%
        {harrison_mediaspace_2009}
\bibfield{editor}{\bibinfo{person}{Steve Harrison}} (Ed.). \bibinfo{year}{2009}\natexlab{}.
\newblock \bibinfo{booktitle}{\emph{Media {Space} 20+ {Years} of {Mediated} {Life}} (\bibinfo{edition}{1st} ed.)}.
\newblock \bibinfo{publisher}{Springer}, \bibinfo{address}{London}.
\newblock
\showISBNx{978-1-84882-482-9}


\bibitem[Haynes et~al\mbox{.}(1997)]%
        {haynes_autoscheduling_1997}
\bibfield{author}{\bibinfo{person}{Thomas Haynes}, \bibinfo{person}{Sandip Sen}, \bibinfo{person}{Neeraj Arora}, {and} \bibinfo{person}{Rajani Nadella}.} \bibinfo{year}{1997}\natexlab{}.
\newblock \showarticletitle{An Automated Meeting Scheduling System That Utilizes User Preferences}. In \bibinfo{booktitle}{\emph{Proceedings of the First International Conference on Autonomous Agents}} (Marina del Rey, California, USA) \emph{(\bibinfo{series}{AGENTS '97})}. \bibinfo{publisher}{Association for Computing Machinery}, \bibinfo{address}{New York, NY, USA}, \bibinfo{pages}{308–315}.
\newblock
\showISBNx{0897918770}
\urldef\tempurl%
\url{https://doi.org/10.1145/267658.267733}
\showDOI{\tempurl}


\bibitem[Heritage(2013)]%
        {heritage2013garfinkel}
\bibfield{author}{\bibinfo{person}{John Heritage}.} \bibinfo{year}{2013}\natexlab{}.
\newblock \bibinfo{booktitle}{\emph{Garfinkel and ethnomethodology}}.
\newblock \bibinfo{publisher}{John Wiley \& Sons}.
\newblock


\bibitem[Holmes(2003)]%
        {holmes2003small}
\bibfield{author}{\bibinfo{person}{Janet Holmes}.} \bibinfo{year}{2003}\natexlab{}.
\newblock \showarticletitle{Small talk at work: Potential problems for workers with an intellectual disability}.
\newblock \bibinfo{journal}{\emph{Research on Language and Social Interaction}} \bibinfo{volume}{36}, \bibinfo{number}{1} (\bibinfo{year}{2003}), \bibinfo{pages}{65--84}.
\newblock


\bibitem[Isenberg et~al\mbox{.}(2012)]%
        {isenberg2012coupling}
\bibfield{author}{\bibinfo{person}{Petra Isenberg}, \bibinfo{person}{Danyel Fisher}, \bibinfo{person}{Sharoda~A. Paul}, \bibinfo{person}{Meredith~Ringel Morris}, \bibinfo{person}{Kori Inkpen}, {and} \bibinfo{person}{Mary Czerwinski}.} \bibinfo{year}{2012}\natexlab{}.
\newblock \showarticletitle{Co-Located Collaborative Visual Analytics around a Tabletop Display}.
\newblock \bibinfo{journal}{\emph{IEEE Transactions on Visualization and Computer Graphics}} \bibinfo{volume}{18}, \bibinfo{number}{5} (\bibinfo{year}{2012}), \bibinfo{pages}{689--702}.
\newblock
\urldef\tempurl%
\url{https://doi.org/10.1109/TVCG.2011.287}
\showDOI{\tempurl}


\bibitem[Jackson et~al\mbox{.}(2016)]%
        {jackson_encouraging_2016}
\bibfield{author}{\bibinfo{person}{Corey Jackson}, \bibinfo{person}{Gabriel Mugar}, \bibinfo{person}{Kevin Crowston}, {and} \bibinfo{person}{Carsten Østerlund}.} \bibinfo{year}{2016}\natexlab{}.
\newblock \showarticletitle{Encouraging {Work} in {Citizen} {Science}: {Experiments} in {Goal} {Setting} and {Anchoring}}. In \bibinfo{booktitle}{\emph{Proceedings of the 19th {ACM} {Conference} on {Computer} {Supported} {Cooperative} {Work} and {Social} {Computing} {Companion}}} \emph{(\bibinfo{series}{{CSCW} '16 {Companion}})}. \bibinfo{publisher}{Association for Computing Machinery}, \bibinfo{address}{New York, NY, USA}, \bibinfo{pages}{297--300}.
\newblock
\showISBNx{978-1-4503-3950-6}
\urldef\tempurl%
\url{https://doi.org/10.1145/2818052.2869129}
\showDOI{\tempurl}


\bibitem[Jerry~Fjermestad(2000)]%
        {hiltz_gss_2015}
\bibfield{author}{\bibinfo{person}{Starr Roxanne~Hiltz Jerry~Fjermestad}.} \bibinfo{year}{2000}\natexlab{}.
\newblock \showarticletitle{Group Support Systems: A Descriptive Evaluation of Case and Field Studies}.
\newblock \bibinfo{journal}{\emph{Journal of Management Information Systems}} \bibinfo{volume}{17}, \bibinfo{number}{3} (\bibinfo{year}{2000}), \bibinfo{pages}{115--159}.
\newblock
\urldef\tempurl%
\url{https://doi.org/10.1080/07421222.2000.11045657}
\showDOI{\tempurl}


\bibitem[Kim and Shah(2016)]%
        {kim2016improving}
\bibfield{author}{\bibinfo{person}{Joseph Kim} {and} \bibinfo{person}{Julie~A Shah}.} \bibinfo{year}{2016}\natexlab{}.
\newblock \showarticletitle{Improving team's consistency of understanding in meetings}.
\newblock \bibinfo{journal}{\emph{IEEE Transactions on Human-Machine Systems}} \bibinfo{volume}{46}, \bibinfo{number}{5} (\bibinfo{year}{2016}), \bibinfo{pages}{625--637}.
\newblock


\bibitem[Kim et~al\mbox{.}(2020)]%
        {kim_meetingbot_2020}
\bibfield{author}{\bibinfo{person}{Soomin Kim}, \bibinfo{person}{Jinsu Eun}, \bibinfo{person}{Changhoon Oh}, \bibinfo{person}{Bongwon Suh}, {and} \bibinfo{person}{Joonhwan Lee}.} \bibinfo{year}{2020}\natexlab{}.
\newblock \showarticletitle{Bot in the Bunch: Facilitating Group Chat Discussion by Improving Efficiency and Participation with a Chatbot}. In \bibinfo{booktitle}{\emph{Proceedings of the 2020 CHI Conference on Human Factors in Computing Systems}} (Honolulu, HI, USA) \emph{(\bibinfo{series}{CHI '20})}. \bibinfo{publisher}{Association for Computing Machinery}, \bibinfo{address}{New York, NY, USA}, \bibinfo{pages}{1–13}.
\newblock
\showISBNx{9781450367080}
\urldef\tempurl%
\url{https://doi.org/10.1145/3313831.3376785}
\showDOI{\tempurl}


\bibitem[Kocielnik et~al\mbox{.}(2018a)]%
        {kocielnik2018designing}
\bibfield{author}{\bibinfo{person}{Rafal Kocielnik}, \bibinfo{person}{Daniel Avrahami}, \bibinfo{person}{Jennifer Marlow}, \bibinfo{person}{Di Lu}, {and} \bibinfo{person}{Gary Hsieh}.} \bibinfo{year}{2018}\natexlab{a}.
\newblock \showarticletitle{Designing for workplace reflection: a chat and voice-based conversational agent}. In \bibinfo{booktitle}{\emph{Proceedings of the 2018 designing interactive systems conference}}. \bibinfo{pages}{881--894}.
\newblock


\bibitem[Kocielnik et~al\mbox{.}(2018b)]%
        {kocielnik_reflecting_2018}
\bibfield{author}{\bibinfo{person}{Rafal Kocielnik}, \bibinfo{person}{Lillian Xiao}, \bibinfo{person}{Daniel Avrahami}, {and} \bibinfo{person}{Gary Hsieh}.} \bibinfo{year}{2018}\natexlab{b}.
\newblock \showarticletitle{Reflection Companion: A Conversational System for Engaging Users in Reflection on Physical Activity}.
\newblock \bibinfo{journal}{\emph{Proc. ACM Interact. Mob. Wearable Ubiquitous Technol.}} \bibinfo{volume}{2}, \bibinfo{number}{2}, Article \bibinfo{articleno}{70} (\bibinfo{date}{jul} \bibinfo{year}{2018}), \bibinfo{numpages}{26}~pages.
\newblock
\urldef\tempurl%
\url{https://doi.org/10.1145/3214273}
\showDOI{\tempurl}


\bibitem[Kuzminykh and Rintel(2020)]%
        {kuzminykh_classification_2020}
\bibfield{author}{\bibinfo{person}{Anastasia Kuzminykh} {and} \bibinfo{person}{Sean Rintel}.} \bibinfo{year}{2020}\natexlab{}.
\newblock \showarticletitle{Classification of {Functional} {Attention} in {Video} {Meetings}}. In \bibinfo{booktitle}{\emph{Proceedings of the 2020 {CHI} {Conference} on {Human} {Factors} in {Computing} {Systems}}} \emph{(\bibinfo{series}{{CHI} '20})}. \bibinfo{publisher}{Association for Computing Machinery}, \bibinfo{address}{New York, NY, USA}, \bibinfo{pages}{1--13}.
\newblock
\showISBNx{978-1-4503-6708-0}
\urldef\tempurl%
\url{https://doi.org/10.1145/3313831.3376546}
\showDOI{\tempurl}


\bibitem[Köhler and Gölz(2015)]%
        {kohler_meetings_2015}
\bibfield{author}{\bibinfo{person}{Tine Köhler} {and} \bibinfo{person}{Markus Gölz}.} \bibinfo{year}{2015}\natexlab{}.
\newblock \showarticletitle{Meetings across cultures: {Cultural} differences in meeting expectations and processes}.
\newblock In \bibinfo{booktitle}{\emph{The {Cambridge} handbook of meeting science}}. \bibinfo{publisher}{Cambridge University Press}, \bibinfo{address}{New York, NY, US}, \bibinfo{pages}{119--149}.
\newblock
\showISBNx{978-1-107-06718-9}
\urldef\tempurl%
\url{https://doi.org/10.1017/CBO9781107589735.007}
\showDOI{\tempurl}


\bibitem[Langley(1995)]%
        {langley1995between}
\bibfield{author}{\bibinfo{person}{Ann Langley}.} \bibinfo{year}{1995}\natexlab{}.
\newblock \showarticletitle{Between'paralysis by analysis' and'extinction by instinct'}.
\newblock \bibinfo{journal}{\emph{MIT Sloan Management Review}} \bibinfo{volume}{36}, \bibinfo{number}{3} (\bibinfo{year}{1995}), \bibinfo{pages}{63}.
\newblock


\bibitem[Latham and Locke(1991)]%
        {latham_self-regulation_1991}
\bibfield{author}{\bibinfo{person}{Gary~P Latham} {and} \bibinfo{person}{Edwin~A Locke}.} \bibinfo{year}{1991}\natexlab{}.
\newblock \showarticletitle{Self-regulation through goal setting}.
\newblock \bibinfo{journal}{\emph{Organizational Behavior and Human Decision Processes}} \bibinfo{volume}{50}, \bibinfo{number}{2} (\bibinfo{date}{Dec.} \bibinfo{year}{1991}), \bibinfo{pages}{212--247}.
\newblock
\showISSN{07495978}
\urldef\tempurl%
\url{https://doi.org/10.1016/0749-5978(91)90021-K}
\showDOI{\tempurl}


\bibitem[Leach et~al\mbox{.}(2009)]%
        {leach_perceived_2009}
\bibfield{author}{\bibinfo{person}{Desmond~J. Leach}, \bibinfo{person}{Steven~G. Rogelberg}, \bibinfo{person}{Peter~B. Warr}, {and} \bibinfo{person}{Jennifer~L. Burnfield}.} \bibinfo{year}{2009}\natexlab{}.
\newblock \showarticletitle{Perceived {Meeting} {Effectiveness}: {The} {Role} of {Design} {Characteristics}}.
\newblock \bibinfo{journal}{\emph{Journal of Business and Psychology}} \bibinfo{volume}{24}, \bibinfo{number}{1} (\bibinfo{date}{March} \bibinfo{year}{2009}), \bibinfo{pages}{65--76}.
\newblock
\showISSN{1573-353X}
\urldef\tempurl%
\url{https://doi.org/10.1007/s10869-009-9092-6}
\showDOI{\tempurl}


\bibitem[Lehmann-Willenbrock and Meinecke(2017)]%
        {lehmann-willenbrock_team-meeting_2017}
\bibfield{author}{\bibinfo{person}{Nale Lehmann-Willenbrock} {and} \bibinfo{person}{Annika~L. Meinecke}.} \bibinfo{year}{2017}\natexlab{}.
\newblock \showarticletitle{Team-{Meeting} {Behaviors} in {Germany} and the {United} {States}}.
\newblock In \bibinfo{booktitle}{\emph{The {International} {Encyclopedia} of {Intercultural} {Communication}}}. \bibinfo{publisher}{John Wiley \& Sons, Ltd}, \bibinfo{pages}{1--6}.
\newblock
\showISBNx{978-1-118-78366-5}
\urldef\tempurl%
\url{https://doi.org/10.1002/9781118783665.ieicc0262}
\showDOI{\tempurl}
\newblock
\shownote{\_eprint: https://onlinelibrary.wiley.com/doi/pdf/10.1002/9781118783665.ieicc0262}.


\bibitem[Lenz and Lyles(1985)]%
        {lenz1985paralysis}
\bibfield{author}{\bibinfo{person}{RT Lenz} {and} \bibinfo{person}{Marjorie~A Lyles}.} \bibinfo{year}{1985}\natexlab{}.
\newblock \showarticletitle{Paralysis by analysis: is your planning system becoming too rational?}
\newblock \bibinfo{journal}{\emph{Long Range Planning}} \bibinfo{volume}{18}, \bibinfo{number}{4} (\bibinfo{year}{1985}), \bibinfo{pages}{64--72}.
\newblock


\bibitem[Li et~al\mbox{.}(2023)]%
        {li_exploring_2023}
\bibfield{author}{\bibinfo{person}{Zhuoyang Li}, \bibinfo{person}{Minhui Liang}, \bibinfo{person}{Hai~Trung Le}, \bibinfo{person}{Ray Lc}, {and} \bibinfo{person}{Yuhan Luo}.} \bibinfo{year}{2023}\natexlab{}.
\newblock \showarticletitle{Exploring {Design} {Opportunities} for {Reflective} {Conversational} {Agents} to {Reduce} {Compulsive} {Smartphone} {Use}}. In \bibinfo{booktitle}{\emph{Proceedings of the 5th {International} {Conference} on {Conversational} {User} {Interfaces}}} \emph{(\bibinfo{series}{{CUI} '23})}. \bibinfo{publisher}{Association for Computing Machinery}, \bibinfo{address}{New York, NY, USA}, \bibinfo{pages}{1--6}.
\newblock
\showISBNx{9798400700149}
\urldef\tempurl%
\url{https://doi.org/10.1145/3571884.3604305}
\showDOI{\tempurl}


\bibitem[Luong and Rogelberg(2005)]%
        {luong_meetings_2005}
\bibfield{author}{\bibinfo{person}{Alexandra Luong} {and} \bibinfo{person}{Steven~G. Rogelberg}.} \bibinfo{year}{2005}\natexlab{}.
\newblock \showarticletitle{Meetings and {More} {Meetings}: {The} {Relationship} {Between} {Meeting} {Load} and the {Daily} {Well}-{Being} of {Employees}}.
\newblock \bibinfo{journal}{\emph{Group Dynamics: Theory, Research, and Practice}} \bibinfo{volume}{9}, \bibinfo{number}{1} (\bibinfo{year}{2005}), \bibinfo{pages}{58--67}.
\newblock
\showISSN{1930-7802}
\urldef\tempurl%
\url{https://doi.org/10.1037/1089-2699.9.1.58}
\showDOI{\tempurl}
\newblock
\shownote{Place: US Publisher: Educational Publishing Foundation}.


\bibitem[López-Fresno and Cascón-Pereira(2022)]%
        {lopez-fresno_what_2022}
\bibfield{author}{\bibinfo{person}{Palmira López-Fresno} {and} \bibinfo{person}{Rosalía Cascón-Pereira}.} \bibinfo{year}{2022}\natexlab{}.
\newblock \showarticletitle{What is the {Purpose} of this {Meeting}? {The} hidden meanings of the meeting announcement}.
\newblock \bibinfo{journal}{\emph{Organization Studies}} \bibinfo{volume}{43}, \bibinfo{number}{8} (\bibinfo{date}{Aug.} \bibinfo{year}{2022}), \bibinfo{pages}{1297--1325}.
\newblock
\showISSN{0170-8406}
\urldef\tempurl%
\url{https://doi.org/10.1177/01708406211040216}
\showDOI{\tempurl}
\newblock
\shownote{Publisher: SAGE Publications Ltd}.


\bibitem[Ma et~al\mbox{.}(2019)]%
        {ma2019interpersonal}
\bibfield{author}{\bibinfo{person}{Jingjing Ma}, \bibinfo{person}{John~M Schaubroeck}, {and} \bibinfo{person}{Catherine LeBlanc}.} \bibinfo{year}{2019}\natexlab{}.
\newblock \showarticletitle{Interpersonal trust in organizations}.
\newblock In \bibinfo{booktitle}{\emph{Oxford Research Encyclopedia of Business and Management}}.
\newblock


\bibitem[McGrath(1991)]%
        {mcgrath1991time}
\bibfield{author}{\bibinfo{person}{Joseph~E McGrath}.} \bibinfo{year}{1991}\natexlab{}.
\newblock \showarticletitle{Time, interaction, and performance (TIP) A Theory of Groups}.
\newblock \bibinfo{journal}{\emph{Small group research}} \bibinfo{volume}{22}, \bibinfo{number}{2} (\bibinfo{year}{1991}), \bibinfo{pages}{147--174}.
\newblock


\bibitem[McParland and Connolly(2020)]%
        {mcparland_dataveillance_2020}
\bibfield{author}{\bibinfo{person}{Cliona McParland} {and} \bibinfo{person}{Regina Connolly}.} \bibinfo{year}{2020}\natexlab{}.
\newblock \showarticletitle{Dataveillance in the {Workplace}: {Managing} the {Impact} of {Innovation}}.
\newblock \bibinfo{journal}{\emph{Business Systems Research : International journal of the Society for Advancing Innovation and Research in Economy}} \bibinfo{volume}{11}, \bibinfo{number}{1} (\bibinfo{date}{June} \bibinfo{year}{2020}), \bibinfo{pages}{106--124}.
\newblock
\showISSN{1847-9375}
\urldef\tempurl%
\url{https://hrcak.srce.hr/ojs/index.php/bsr/article/view/12670}
\showURL{%
\tempurl}
\newblock
\shownote{Number: 1}.


\bibitem[Meyer et~al\mbox{.}(2021)]%
        {meyer_enabling_2021}
\bibfield{author}{\bibinfo{person}{André~N. Meyer}, \bibinfo{person}{Gail~C. Murphy}, \bibinfo{person}{Thomas Zimmermann}, {and} \bibinfo{person}{Thomas Fritz}.} \bibinfo{year}{2021}\natexlab{}.
\newblock \showarticletitle{Enabling {Good} {Work} {Habits} in {Software} {Developers} through {Reflective} {Goal}-{Setting}}.
\newblock \bibinfo{journal}{\emph{IEEE Transactions on Software Engineering}} \bibinfo{volume}{47}, \bibinfo{number}{9} (\bibinfo{date}{Sept.} \bibinfo{year}{2021}), \bibinfo{pages}{1872--1885}.
\newblock
\showISSN{1939-3520}
\urldef\tempurl%
\url{https://doi.org/10.1109/TSE.2019.2938525}
\showDOI{\tempurl}
\newblock
\shownote{Conference Name: IEEE Transactions on Software Engineering}.


\bibitem[{Microsoft}(2023)]%
        {microsoft_work_2023}
\bibfield{author}{\bibinfo{person}{{Microsoft}}.} \bibinfo{year}{2023}\natexlab{}.
\newblock \bibinfo{title}{Work {Trend} {Index} {\textbar} {Will} {AI} {Fix} {Work}?}
\newblock
\newblock
\urldef\tempurl%
\url{https://www.microsoft.com/en-us/worklab/will-ai-fix-work}
\showURL{%
\tempurl}


\bibitem[Morshed et~al\mbox{.}(2022)]%
        {mehrab-czerwinski_stressatwork_2022}
\bibfield{author}{\bibinfo{person}{Mehrab~Bin Morshed}, \bibinfo{person}{Javier Hernandez}, \bibinfo{person}{Daniel McDuff}, \bibinfo{person}{Jina Suh}, \bibinfo{person}{Esther Howe}, \bibinfo{person}{Kael Rowan}, \bibinfo{person}{Marah Abdin}, \bibinfo{person}{Gonzalo Ramos}, \bibinfo{person}{Tracy Tran}, {and} \bibinfo{person}{Mary Czerwinski}.} \bibinfo{year}{2022}\natexlab{}.
\newblock \showarticletitle{Advancing the Understanding and Measurement of Workplace Stress in Remote Information Workers from Passive Sensors and Behavioral Data}. In \bibinfo{booktitle}{\emph{2022 10th International Conference on Affective Computing and Intelligent Interaction (ACII)}}. \bibinfo{pages}{1--8}.
\newblock
\urldef\tempurl%
\url{https://doi.org/10.1109/ACII55700.2022.9953824}
\showDOI{\tempurl}


\bibitem[Neumayr et~al\mbox{.}(2021)]%
        {neumayr2021hybrid}
\bibfield{author}{\bibinfo{person}{Thomas Neumayr}, \bibinfo{person}{Banu Saatci}, \bibinfo{person}{Sean Rintel}, \bibinfo{person}{Clemens~Nylandsted Klokmose}, {and} \bibinfo{person}{Mirjam Augstein}.} \bibinfo{year}{2021}\natexlab{}.
\newblock \showarticletitle{What was hybrid? a systematic review of hybrid collaboration and meetings research}.
\newblock \bibinfo{journal}{\emph{arXiv preprint arXiv:2111.06172}} (\bibinfo{year}{2021}).
\newblock


\bibitem[Niemantsverdriet and Erickson(2017)]%
        {niemantsverdriet_recurring_2017}
\bibfield{author}{\bibinfo{person}{Karin Niemantsverdriet} {and} \bibinfo{person}{Thomas Erickson}.} \bibinfo{year}{2017}\natexlab{}.
\newblock \showarticletitle{Recurring {Meetings}: {An} {Experiential} {Account} of {Repeating} {Meetings} in a {Large} {Organization}}.
\newblock \bibinfo{journal}{\emph{Proceedings of the ACM on Human-Computer Interaction}} \bibinfo{volume}{1}, \bibinfo{number}{CSCW} (\bibinfo{date}{Dec.} \bibinfo{year}{2017}), \bibinfo{pages}{84:1--84:17}.
\newblock
\urldef\tempurl%
\url{https://doi.org/10.1145/3134719}
\showDOI{\tempurl}


\bibitem[Nixon and Littlepage(1992)]%
        {nixon_impact_1992}
\bibfield{author}{\bibinfo{person}{Carol~T. Nixon} {and} \bibinfo{person}{Glenn~E. Littlepage}.} \bibinfo{year}{1992}\natexlab{}.
\newblock \showarticletitle{Impact of meeting procedures on meeting effectiveness}.
\newblock \bibinfo{journal}{\emph{Journal of Business and Psychology}} \bibinfo{volume}{6}, \bibinfo{number}{3} (\bibinfo{year}{1992}), \bibinfo{pages}{361--369}.
\newblock
\showISSN{0889-3268, 1573-353X}
\urldef\tempurl%
\url{https://doi.org/10.1007/BF01126771}
\showDOI{\tempurl}


\bibitem[Norman(1993)]%
        {norman_things_1993}
\bibfield{author}{\bibinfo{person}{Donald~A. Norman}.} \bibinfo{year}{1993}\natexlab{}.
\newblock \bibinfo{booktitle}{\emph{Things that {Make} {Us} {Smart}: {Defending} {Human} {Attributes} in the {Age} of the {Machine}}}.
\newblock \bibinfo{publisher}{Addison-Wesley Publishing Company}.
\newblock
\showISBNx{978-0-201-58129-4}
\newblock
\shownote{Google-Books-ID: Gr9fQgAACAAJ}.


\bibitem[Nunamaker et~al\mbox{.}(1991)]%
        {nunamaker_electronic_1991}
\bibfield{author}{\bibinfo{person}{Jay~F. Nunamaker}, \bibinfo{person}{Alan~R. Dennis}, \bibinfo{person}{Joseph~S. Valacich}, \bibinfo{person}{Douglas Vogel}, {and} \bibinfo{person}{Joey~F. George}.} \bibinfo{year}{1991}\natexlab{}.
\newblock \showarticletitle{Electronic meeting systems}.
\newblock \bibinfo{journal}{\emph{Commun. ACM}} \bibinfo{volume}{34}, \bibinfo{number}{7} (\bibinfo{year}{1991}), \bibinfo{pages}{40--61}.
\newblock
\newblock
\shownote{Publisher: ACM New York, NY, USA}.


\bibitem[Oh et~al\mbox{.}(2018)]%
        {oh2018systematic}
\bibfield{author}{\bibinfo{person}{Catherine~S Oh}, \bibinfo{person}{Jeremy~N Bailenson}, {and} \bibinfo{person}{Gregory~F Welch}.} \bibinfo{year}{2018}\natexlab{}.
\newblock \showarticletitle{A systematic review of social presence: Definition, antecedents, and implications}.
\newblock \bibinfo{journal}{\emph{Frontiers in Robotics and AI}}  \bibinfo{volume}{5} (\bibinfo{year}{2018}), \bibinfo{pages}{409295}.
\newblock


\bibitem[Raikundalia(1998)]%
        {raikundalia_autoagenda_1998}
\bibfield{author}{\bibinfo{person}{G.K. Raikundalia}.} \bibinfo{year}{1998}\natexlab{}.
\newblock \showarticletitle{A Web tool for asynchronous, collaborative development of electronic meeting agendas}. In \bibinfo{booktitle}{\emph{Proceedings. 3rd Asia Pacific Computer Human Interaction (Cat. No.98EX110)}}. \bibinfo{pages}{374--379}.
\newblock
\urldef\tempurl%
\url{https://doi.org/10.1109/APCHI.1998.704464}
\showDOI{\tempurl}


\bibitem[Reicherts et~al\mbox{.}(2022)]%
        {reicherts_extending_2022}
\bibfield{author}{\bibinfo{person}{Leon Reicherts}, \bibinfo{person}{Gun~Woo Park}, {and} \bibinfo{person}{Yvonne Rogers}.} \bibinfo{year}{2022}\natexlab{}.
\newblock \showarticletitle{Extending {Chatbots} to {Probe} {Users}: {Enhancing} {Complex} {Decision}-{Making} {Through} {Probing} {Conversations}}. In \bibinfo{booktitle}{\emph{Proceedings of the 4th {Conference} on {Conversational} {User} {Interfaces}}} \emph{(\bibinfo{series}{{CUI} '22})}. \bibinfo{publisher}{Association for Computing Machinery}, \bibinfo{address}{New York, NY, USA}, \bibinfo{pages}{1--10}.
\newblock
\showISBNx{978-1-4503-9739-1}
\urldef\tempurl%
\url{https://doi.org/10.1145/3543829.3543832}
\showDOI{\tempurl}


\bibitem[Reicherts and Rogers(2020)]%
        {reicherts_make_2020}
\bibfield{author}{\bibinfo{person}{Leon Reicherts} {and} \bibinfo{person}{Yvonne Rogers}.} \bibinfo{year}{2020}\natexlab{}.
\newblock \showarticletitle{Do {Make} me {Think}!: {How} {CUIs} {Can} {Support} {Cognitive} {Processes}}. In \bibinfo{booktitle}{\emph{Proceedings of the 2nd {Conference} on {Conversational} {User} {Interfaces}}}. \bibinfo{publisher}{ACM}, \bibinfo{address}{Bilbao Spain}, \bibinfo{pages}{1--4}.
\newblock
\showISBNx{978-1-4503-7544-3}
\urldef\tempurl%
\url{https://doi.org/10.1145/3405755.3406157}
\showDOI{\tempurl}


\bibitem[Rheingold(1985)]%
        {rheingold_tools_1985}
\bibfield{author}{\bibinfo{person}{Howard Rheingold}.} \bibinfo{year}{1985}\natexlab{}.
\newblock \bibinfo{booktitle}{\emph{Tools for {Thought}: {The} {People} and {Ideas} behind the {Next} {Computer} {Revolution}}}.
\newblock \bibinfo{publisher}{Simon \& Schuster Trade}.
\newblock
\showISBNx{978-0-671-49292-2}


\bibitem[Rogelberg(2018)]%
        {rogelberg2018surprising}
\bibfield{author}{\bibinfo{person}{Steven~G Rogelberg}.} \bibinfo{year}{2018}\natexlab{}.
\newblock \bibinfo{booktitle}{\emph{The surprising science of meetings: How you can lead your team to peak performance}}.
\newblock \bibinfo{publisher}{Oxford University Press, USA}.
\newblock


\bibitem[Rogelberg et~al\mbox{.}(2010)]%
        {rogelberg2010employee}
\bibfield{author}{\bibinfo{person}{Steven~G Rogelberg}, \bibinfo{person}{Joseph~A Allen}, \bibinfo{person}{Linda Shanock}, \bibinfo{person}{Cliff Scott}, {and} \bibinfo{person}{Marissa Shuffler}.} \bibinfo{year}{2010}\natexlab{}.
\newblock \showarticletitle{Employee satisfaction with meetings: A contemporary facet of job satisfaction}.
\newblock \bibinfo{journal}{\emph{Human Resource Management: Published in Cooperation with the School of Business Administration, The University of Michigan and in alliance with the Society of Human Resources Management}} \bibinfo{volume}{49}, \bibinfo{number}{2} (\bibinfo{year}{2010}), \bibinfo{pages}{149--172}.
\newblock


\bibitem[Rogelberg et~al\mbox{.}(2006)]%
        {rogelberg_not_2006}
\bibfield{author}{\bibinfo{person}{Steven~G. Rogelberg}, \bibinfo{person}{Desmond~J. Leach}, \bibinfo{person}{Peter~B. Warr}, {and} \bibinfo{person}{Jennifer~L. Burnfield}.} \bibinfo{year}{2006}\natexlab{}.
\newblock \showarticletitle{"{Not} {Another} {Meeting}!" {Are} {Meeting} {Time} {Demands} {Related} to {Employee} {Well}-{Being}?}
\newblock \bibinfo{journal}{\emph{Journal of Applied Psychology}} \bibinfo{volume}{91}, \bibinfo{number}{1} (\bibinfo{year}{2006}), \bibinfo{pages}{83--96}.
\newblock
\showISSN{1939-1854, 0021-9010}
\urldef\tempurl%
\url{https://doi.org/10.1037/0021-9010.91.1.83}
\showDOI{\tempurl}


\bibitem[Romano and Nunamaker(2001)]%
        {romano_meeting_2001}
\bibfield{author}{\bibinfo{person}{N.C. Romano} {and} \bibinfo{person}{J.F. Nunamaker}.} \bibinfo{year}{2001}\natexlab{}.
\newblock \showarticletitle{Meeting analysis: findings from research and practice}. In \bibinfo{booktitle}{\emph{Proceedings of the 34th {Annual} {Hawaii} {International} {Conference} on {System} {Sciences}}}. \bibinfo{pages}{13 pp.--}.
\newblock
\urldef\tempurl%
\url{https://doi.org/10.1109/HICSS.2001.926253}
\showDOI{\tempurl}


\bibitem[Samrose et~al\mbox{.}(2021)]%
        {samrose_meetingcoach_2021}
\bibfield{author}{\bibinfo{person}{Samiha Samrose}, \bibinfo{person}{Daniel McDuff}, \bibinfo{person}{Robert Sim}, \bibinfo{person}{Jina Suh}, \bibinfo{person}{Kael Rowan}, \bibinfo{person}{Javier Hernandez}, \bibinfo{person}{Sean Rintel}, \bibinfo{person}{Kevin Moynihan}, {and} \bibinfo{person}{Mary Czerwinski}.} \bibinfo{year}{2021}\natexlab{}.
\newblock \showarticletitle{{MeetingCoach}: {An} {Intelligent} {Dashboard} for {Supporting} {Effective} \& {Inclusive} {Meetings}}. In \bibinfo{booktitle}{\emph{Proceedings of the 2021 {CHI} {Conference} on {Human} {Factors} in {Computing} {Systems}}} \emph{(\bibinfo{series}{{CHI} '21})}. \bibinfo{publisher}{Association for Computing Machinery}, \bibinfo{address}{New York, NY, USA}, \bibinfo{pages}{1--13}.
\newblock
\showISBNx{978-1-4503-8096-6}
\urldef\tempurl%
\url{https://doi.org/10.1145/3411764.3445615}
\showDOI{\tempurl}


\bibitem[Sarkar et~al\mbox{.}(2021)]%
        {sarkar_promise_2021}
\bibfield{author}{\bibinfo{person}{Advait Sarkar}, \bibinfo{person}{Sean Rintel}, \bibinfo{person}{Damian Borowiec}, \bibinfo{person}{Rachel Bergmann}, \bibinfo{person}{Sharon Gillett}, \bibinfo{person}{Danielle Bragg}, \bibinfo{person}{Nancy Baym}, {and} \bibinfo{person}{Abigail Sellen}.} \bibinfo{year}{2021}\natexlab{}.
\newblock \showarticletitle{The promise and peril of parallel chat in video meetings for work}. In \bibinfo{booktitle}{\emph{Extended {Abstracts} of the 2021 {CHI} {Conference} on {Human} {Factors} in {Computing} {Systems}}} \emph{(\bibinfo{series}{{CHI} {EA} '21})}. \bibinfo{publisher}{Association for Computing Machinery}, \bibinfo{address}{New York, NY, USA}, \bibinfo{pages}{1--8}.
\newblock
\showISBNx{978-1-4503-8095-9}
\urldef\tempurl%
\url{https://doi.org/10.1145/3411763.3451793}
\showDOI{\tempurl}


\bibitem[Schmidt(2013)]%
        {schmidt_cscwfoundations_2013}
\bibfield{author}{\bibinfo{person}{Kjeld Schmidt}.} \bibinfo{year}{2013}\natexlab{}.
\newblock \bibinfo{booktitle}{\emph{Cooperative Work and Coordinative Practices: Contributions to the Conceptual Foundations of Computer-Supported Cooperative Work (CSCW)}}.
\newblock \bibinfo{publisher}{Springer Publishing Company, Incorporated}.
\newblock
\showISBNx{1447126319}


\bibitem[Scott et~al\mbox{.}(2015)]%
        {scott2015five}
\bibfield{author}{\bibinfo{person}{Cliff Scott}, \bibinfo{person}{Joseph~A Allen}, \bibinfo{person}{Steven~G Rogelberg}, {and} \bibinfo{person}{Alex Kello}.} \bibinfo{year}{2015}\natexlab{}.
\newblock \showarticletitle{Five theoretical lenses for conceptualizing the role of meetings in organizational life}. In \bibinfo{booktitle}{\emph{The Cambridge handbook of meeting science}}. Cambridge University Press New York, NY, \bibinfo{pages}{20--46}.
\newblock


\bibitem[Scott and Wildman(2015)]%
        {scott2015culture}
\bibfield{author}{\bibinfo{person}{Charles~PR Scott} {and} \bibinfo{person}{Jessica~L Wildman}.} \bibinfo{year}{2015}\natexlab{}.
\newblock \showarticletitle{Culture, communication, and conflict: A review of the global virtual team literature}.
\newblock \bibinfo{journal}{\emph{Leading global teams: Translating multidisciplinary science to practice}} (\bibinfo{year}{2015}), \bibinfo{pages}{13--32}.
\newblock


\bibitem[Sheeran and Webb(2016)]%
        {sheeran_intention-behavior_2016}
\bibfield{author}{\bibinfo{person}{Paschal Sheeran} {and} \bibinfo{person}{Thomas~L. Webb}.} \bibinfo{year}{2016}\natexlab{}.
\newblock \showarticletitle{The {Intention}-{Behavior} {Gap}: {The} {Intention}-{Behavior} {Gap}}.
\newblock \bibinfo{journal}{\emph{Social and Personality Psychology Compass}} \bibinfo{volume}{10}, \bibinfo{number}{9} (\bibinfo{date}{Sept.} \bibinfo{year}{2016}), \bibinfo{pages}{503--518}.
\newblock
\showISSN{17519004}
\urldef\tempurl%
\url{https://doi.org/10.1111/spc3.12265}
\showDOI{\tempurl}


\bibitem[Soria et~al\mbox{.}(2022)]%
        {soria_recurring_2022}
\bibfield{author}{\bibinfo{person}{Adriana~Meza Soria}, \bibinfo{person}{André van~der Hoek}, {and} \bibinfo{person}{Janet Burge}.} \bibinfo{year}{2022}\natexlab{}.
\newblock \showarticletitle{Recurring distributed software maintenance meetings: toward an initial understanding}. In \bibinfo{booktitle}{\emph{Proceedings of the 15th {International} {Conference} on {Cooperative} and {Human} {Aspects} of {Software} {Engineering}}} \emph{(\bibinfo{series}{{CHASE} '22})}. \bibinfo{publisher}{Association for Computing Machinery}, \bibinfo{address}{New York, NY, USA}, \bibinfo{pages}{21--25}.
\newblock
\showISBNx{978-1-4503-9342-3}
\urldef\tempurl%
\url{https://doi.org/10.1145/3528579.3529179}
\showDOI{\tempurl}


\bibitem[Standaert et~al\mbox{.}(2016)]%
        {standaert_empirical_2016}
\bibfield{author}{\bibinfo{person}{Willem Standaert}, \bibinfo{person}{Steve Muylle}, {and} \bibinfo{person}{Amit Basu}.} \bibinfo{year}{2016}\natexlab{}.
\newblock \showarticletitle{An {Empirical} {Study} of the {Effectiveness} of {Telepresence} as a {Business} {Meeting} {Mode}}.
\newblock \bibinfo{journal}{\emph{Information Technology and Management}}  \bibinfo{volume}{17} (\bibinfo{date}{Dec.} \bibinfo{year}{2016}), \bibinfo{pages}{323--339}.
\newblock
\urldef\tempurl%
\url{https://doi.org/10.1007/s10799-015-0221-9}
\showDOI{\tempurl}


\bibitem[Standaert et~al\mbox{.}(2021)]%
        {standaert_how_2021}
\bibfield{author}{\bibinfo{person}{Willem Standaert}, \bibinfo{person}{Steve Muylle}, {and} \bibinfo{person}{Amit Basu}.} \bibinfo{year}{2021}\natexlab{}.
\newblock \showarticletitle{How shall we meet? {Understanding} the importance of meeting mode capabilities for different meeting objectives}.
\newblock \bibinfo{journal}{\emph{Information \& Management}} \bibinfo{volume}{58}, \bibinfo{number}{1} (\bibinfo{date}{Jan.} \bibinfo{year}{2021}), \bibinfo{pages}{103393}.
\newblock
\showISSN{0378-7206}
\urldef\tempurl%
\url{https://doi.org/10.1016/j.im.2020.103393}
\showDOI{\tempurl}


\bibitem[Strauss and Corbin(2004)]%
        {strauss2004open}
\bibfield{author}{\bibinfo{person}{Anselm~L Strauss} {and} \bibinfo{person}{Juliet Corbin}.} \bibinfo{year}{2004}\natexlab{}.
\newblock \showarticletitle{Open coding}.
\newblock \bibinfo{journal}{\emph{Social research methods: A reader}} (\bibinfo{year}{2004}), \bibinfo{pages}{303--306}.
\newblock


\bibitem[Sun et~al\mbox{.}(2023)]%
        {sun_rhythm_2023}
\bibfield{author}{\bibinfo{person}{Lu Sun}, \bibinfo{person}{Lillio Mok}, \bibinfo{person}{Shilad Sen}, {and} \bibinfo{person}{Bahar Sarrafzadeh}.} \bibinfo{year}{2023}\natexlab{}.
\newblock \bibinfo{title}{Rhythm of {Work}: {Mixed}-methods {Characterization} of {Information} {Workers} {Scheduling} {Preferences} and {Practices}}.
\newblock
\newblock
\urldef\tempurl%
\url{http://arxiv.org/abs/2309.08104}
\showURL{%
\tempurl}
\newblock
\shownote{arXiv:2309.08104 [cs] version: 1}.


\bibitem[Tang et~al\mbox{.}(2023)]%
        {tang2023hybrid}
\bibfield{author}{\bibinfo{person}{John~C. Tang}, \bibinfo{person}{Kori Inkpen}, \bibinfo{person}{Sasa Junuzovic}, \bibinfo{person}{Keri Mallari}, \bibinfo{person}{Andrew~D. Wilson}, \bibinfo{person}{Sean Rintel}, \bibinfo{person}{Shiraz Cupala}, \bibinfo{person}{Tony Carbary}, \bibinfo{person}{Abigail Sellen}, {and} \bibinfo{person}{William~A.S. Buxton}.} \bibinfo{year}{2023}\natexlab{}.
\newblock \showarticletitle{Perspectives: Creating Inclusive and Equitable Hybrid Meeting Experiences}.
\newblock \bibinfo{journal}{\emph{Proc. ACM Hum.-Comput. Interact.}} \bibinfo{volume}{7}, \bibinfo{number}{CSCW2}, Article \bibinfo{articleno}{351} (\bibinfo{date}{oct} \bibinfo{year}{2023}), \bibinfo{numpages}{25}~pages.
\newblock
\urldef\tempurl%
\url{https://doi.org/10.1145/3610200}
\showDOI{\tempurl}


\bibitem[Teevan et~al\mbox{.}(2020)]%
        {teevan2020new}
\bibfield{author}{\bibinfo{person}{Jaime Teevan}, \bibinfo{person}{B Hecht}, {and} \bibinfo{person}{S Jaffe}.} \bibinfo{year}{2020}\natexlab{}.
\newblock \showarticletitle{The new future of work}.
\newblock \bibinfo{journal}{\emph{Microsoft Internal Report}} (\bibinfo{year}{2020}).
\newblock


\bibitem[Trapnell and Campbell(1999)]%
        {trapnell1999private}
\bibfield{author}{\bibinfo{person}{Paul~D Trapnell} {and} \bibinfo{person}{Jennifer~D Campbell}.} \bibinfo{year}{1999}\natexlab{}.
\newblock \showarticletitle{Private self-consciousness and the five-factor model of personality: distinguishing rumination from reflection.}
\newblock \bibinfo{journal}{\emph{Journal of personality and social psychology}} \bibinfo{volume}{76}, \bibinfo{number}{2} (\bibinfo{year}{1999}), \bibinfo{pages}{284}.
\newblock


\bibitem[Tropman(2003)]%
        {tropman2003making}
\bibfield{author}{\bibinfo{person}{John~E Tropman}.} \bibinfo{year}{2003}\natexlab{}.
\newblock \bibinfo{booktitle}{\emph{Making meetings work: Achieving high quality group decisions}}.
\newblock \bibinfo{publisher}{Sage}.
\newblock


\bibitem[Turoff et~al\mbox{.}(1993)]%
        {turoff_gss_1993}
\bibfield{author}{\bibinfo{person}{Murray Turoff}, \bibinfo{person}{Starr~Roxanne Hiltz}, \bibinfo{person}{Ahmed N.~F. Bahgat}, {and} \bibinfo{person}{Ajaz~R. Rana}.} \bibinfo{year}{1993}\natexlab{}.
\newblock \showarticletitle{Distributed Group Support Systems}.
\newblock \bibinfo{journal}{\emph{MIS Quarterly}} \bibinfo{volume}{17}, \bibinfo{number}{4} (\bibinfo{year}{1993}), \bibinfo{pages}{399--417}.
\newblock
\showISSN{02767783}
\urldef\tempurl%
\url{http://www.jstor.org/stable/249585}
\showURL{%
\tempurl}


\bibitem[van Eerde and Buengeler(2015)]%
        {van_eerde_meetings_2015}
\bibfield{author}{\bibinfo{person}{Wendelien van Eerde} {and} \bibinfo{person}{Claudia Buengeler}.} \bibinfo{year}{2015}\natexlab{}.
\newblock \showarticletitle{Meetings all over the world: {Structural} and psychological characteristics of meetings in different countries}.
\newblock In \bibinfo{booktitle}{\emph{The {Cambridge} handbook of meeting science}}. \bibinfo{publisher}{Cambridge University Press}, \bibinfo{address}{New York, NY, US}, \bibinfo{pages}{177--202}.
\newblock
\showISBNx{978-1-107-06718-9}
\urldef\tempurl%
\url{https://doi.org/10.1017/CBO9781107589735.009}
\showDOI{\tempurl}


\bibitem[van Vree(2019)]%
        {vanvree_formalisationofmeetings_2019}
\bibfield{author}{\bibinfo{person}{Wilbert van Vree}.} \bibinfo{year}{2019}\natexlab{}.
\newblock \showarticletitle{Formalisation and {Informalisation} of {Meeting} {Manners}}.
\newblock In \bibinfo{booktitle}{\emph{Civilisation and {Informalisation}: {Connecting} {Long}-{Term} {Social} and {Psychic} {Processes}}}, \bibfield{editor}{\bibinfo{person}{Cas Wouters} {and} \bibinfo{person}{Michael Dunning}} (Eds.). \bibinfo{publisher}{Springer International Publishing}, \bibinfo{address}{Cham}, \bibinfo{pages}{291--313}.
\newblock
\showISBNx{978-3-030-00798-0}
\urldef\tempurl%
\url{https://doi.org/10.1007/978-3-030-00798-0_11}
\showDOI{\tempurl}


\bibitem[Yeomans et~al\mbox{.}(2022)]%
        {yeomans_conversational_2022}
\bibfield{author}{\bibinfo{person}{Michael Yeomans}, \bibinfo{person}{Maurice~E. Schweitzer}, {and} \bibinfo{person}{Alison~Wood Brooks}.} \bibinfo{year}{2022}\natexlab{}.
\newblock \showarticletitle{The {Conversational} {Circumplex}: {Identifying}, prioritizing, and pursuing informational and relational motives in conversation}.
\newblock \bibinfo{journal}{\emph{Current Opinion in Psychology}}  \bibinfo{volume}{44} (\bibinfo{date}{April} \bibinfo{year}{2022}), \bibinfo{pages}{293--302}.
\newblock
\showISSN{2352-250X}
\urldef\tempurl%
\url{https://doi.org/10.1016/j.copsyc.2021.10.001}
\showDOI{\tempurl}


\bibitem[Yoerger et~al\mbox{.}(2015)]%
        {yoerger2015participate}
\bibfield{author}{\bibinfo{person}{Michael Yoerger}, \bibinfo{person}{John Crowe}, {and} \bibinfo{person}{Joseph~A Allen}.} \bibinfo{year}{2015}\natexlab{}.
\newblock \showarticletitle{Participate or else!: The effect of participation in decision-making in meetings on employee engagement.}
\newblock \bibinfo{journal}{\emph{Consulting Psychology Journal: Practice and Research}} \bibinfo{volume}{67}, \bibinfo{number}{1} (\bibinfo{year}{2015}), \bibinfo{pages}{65}.
\newblock


\bibitem[Zhou et~al\mbox{.}(2021a)]%
        {zhou2021role}
\bibfield{author}{\bibinfo{person}{Ke Zhou}, \bibinfo{person}{Marios Constantinides}, \bibinfo{person}{Luca~Maria Aiello}, \bibinfo{person}{Sagar Joglekar}, {and} \bibinfo{person}{Daniele Quercia}.} \bibinfo{year}{2021}\natexlab{a}.
\newblock \showarticletitle{The Role of Different Types of Conversations for Meeting Success}.
\newblock \bibinfo{journal}{\emph{IEEE Pervasive Computing}} \bibinfo{volume}{20}, \bibinfo{number}{4} (\bibinfo{year}{2021}), \bibinfo{pages}{35--42}.
\newblock


\bibitem[Zhou et~al\mbox{.}(2021b)]%
        {zhou_navigating_2023}
\bibfield{author}{\bibinfo{person}{Ke Zhou}, \bibinfo{person}{Marios Constantinides}, \bibinfo{person}{Luca~Maria Aiello}, \bibinfo{person}{Sagar Joglekar}, {and} \bibinfo{person}{Daniele Quercia}.} \bibinfo{year}{2021}\natexlab{b}.
\newblock \showarticletitle{The Role of Different Types of Conversations for Meeting Success}.
\newblock \bibinfo{journal}{\emph{IEEE Pervasive Computing}} \bibinfo{volume}{20}, \bibinfo{number}{4} (\bibinfo{year}{2021}), \bibinfo{pages}{35--42}.
\newblock
\urldef\tempurl%
\url{https://doi.org/10.1109/MPRV.2021.3115879}
\showDOI{\tempurl}


\end{thebibliography}
\appendix

\end{document}